\documentclass[fleqn,usenatbib]{mnras}

\usepackage{newtxtext,newtxmath}

\usepackage[T1]{fontenc}

\DeclareRobustCommand{\VAN}[3]{#2}
\let\VANthebibliography\thebibliography
\def\thebibliography{\DeclareRobustCommand{\VAN}[3]{##3}\VANthebibliography}

\usepackage{graphicx}	
\usepackage{amsmath}	
\usepackage[usenames,dvipsnames]{color}

\title[Efficacy of Radio-Mechanical Feedback]{Molecular Flows in Contemporary Active Galaxies and the Efficacy of Radio-Mechanical Feedback}

\author[P. D. Tamhane et al.]{
Prathamesh D. Tamhane,$^{1,2}$\thanks{E-mail: pdtamhane@uwaterloo.ca}
Brian R. McNamara,$^{1,2,3}$\thanks{E-mail: mcnamara@uwaterloo.ca}
Helen R. Russell,$^{4}$
Alastair C. Edge,$^{5}$
\newauthor
Andrew C. Fabian,$^{6}$
Paul E. J. Nulsen,$^{7,8}$
and Iurii V. Babyk$^{9,10}$
\\
$^{1}$Department of Physics and Astronomy, University of Waterloo, Waterloo, ON N2L 3G1, Canada\\
$^{2}$Waterloo Centre for Astrophysics, University of Waterloo, Waterloo, ON, N2L 3G1, Canada\\
$^{3}$Perimeter Institute for Theoretical Physics, 31 Caroline Street N, Waterloo, ON N2L 2Y5, Canada\\
$^{4}$Department of Physics and Astronomy, University of Nottingham, University Park, Nottingham NG7 2RD, UK\\
$^{5}$Department of Physics, Durham University, Durham DH1 3LE, UK\\
$^{6}$Institute of Astronomy, Madingley Road, Cambridge CB3 0HA, UK\\
$^{7}$Harvard-Smithsonian Center for Astrophysics, 60 Garden Street, Cambridge, MA 02138, USA\\
$^{8}$ICRAR, University of Western Australia, 35 Stirling Hwy, Crawley, WA 6009, Australia\\
$^{9}$Department of Physics and Astronomy, University of California, Irvine, 4129 Frederick Reines Hall, Irvine, CA 92697-4575, USA\\
$^{10}$Main Astronomical Observatory of the National Academy of Sciences of Ukraine, 27 Academica Zabolotnoho str., 03143, Kyiv, Ukraine
}

\date{Accepted XXX. Received YYY; in original form ZZZ}

\pubyear{2022}

\begin{document}
\label{firstpage}
\pagerange{\pageref{firstpage}--\pageref{lastpage}}
\maketitle

\begin{abstract}
Molecular gas flows are analyzed in 14 cluster galaxies (BCGs) centered in cooling hot atmospheres. The BCGs contain $10^{9}-10^{11}~\rm M_\odot$ of molecular gas, much of which is being moved by radio jets and lobes. The molecular flows and radio jet powers are compared to molecular outflows in 45 active galaxies within $z<0.2$. We seek to understand the relative efficacy of radio, quasar, and starburst feedback over a range of active galaxy types.  Molecular flows powered by radio feedback in BCGs are $\sim$10--1000 times larger in extent compared to contemporary galaxies hosting quasar nuclei and starbursts.  Radio feedback yields lower flow velocities but higher momenta compared to quasar nuclei,  as the molecular gas flows in BCGs are usually $\sim$10--100 times more massive. 
The product of the molecular gas mass and lifting altitude divided by the AGN or starburst power --- a parameter referred to as the lifting factor---exceeds starbursts and quasar nuclei by two to three orders of magnitude, respectively.
When active, radio feedback is generally more effective at lifting gas in galaxies compared to quasars and starburst winds. The kinetic energy flux of molecular clouds generally lies below and often substantially below a few percent of the driving power. We find tentatively that star formation is suppressed in BCGs relative to other active galaxies, perhaps because these systems rarely form molecular disks that are more impervious to feedback and are better able to promote star formation.
\end{abstract}




{\section{Introduction}}
	\label{sec:introduction}
    The energy released by active galactic nuclei (AGN) and star formation is able to lift and perhaps expel interstellar gas (ISM) in galaxies at all epochs. 
    Gaseous outflows of ionized, atomic, and molecular gas delay and suppress star formation in galaxies, affecting their evolutionary paths \citep{alatalo11,maiolino12,aalto12a,cicone12,combes13,veilleux13,tombesi15,feruglio15,rupke17}. For example, energetic feedback from starburst winds and AGN may imprint the correlation between stellar velocity dispersion of the host galaxy bulges and the supermassive black holes (SMBHs) masses  \citep{fabian12,king15,croton08,bower08} and may prevent galaxies from growing to much larger sizes \citep{scannapieco04,beckmann17}.  Burgeoning galaxies at redshift 2 are maintained on the main sequence of galaxy formation by in- and outflowing gas propelled by star formation and AGN \citep{tacconi20}.

	Observations have established that outflows from galaxies that host powerful AGN are primarily powered by the AGN \citep{cicone14}. In the most massive galaxies, two primary forms or modes of AGN feedback are thought to operate during a galaxy's lifetime: the radio and quasar modes \citep{bower06,croton06}. During the quasar mode, nuclear winds drive large-scale outflows of gas \citep{tombesi15}.   
	These winds are thought to eject interstellar gas from the nascent galaxy which suppresses star formation and accretion onto the central SMBH \citep{fabian12,veilleux20}. The mode, quasar or radio, is thought to be governed by the specific accretion rate onto the black hole \citep{churazov05}. For accretion rates lower than a few percent of the Eddington rate, radio-mechanical feedback ensues \citep{russell13}. Accretion rates approaching the Eddington rate produce a quasar.  
	
	The $M-\sigma$ relation is thought to have been imprinted by quasars accreting near the Eddington rate following major mergers \citep{kauffmann00}. Whether an Eddington flow conserves energy or momentum is consequential to the form and development of the $M-\sigma$ relation. A hot, adiabatic wind whose radiative cooling timescale exceeds the outflow timescale conserves energy.  Energy conserving winds lead to $M-\sigma$ relations scaling as $\sigma^5$ \citep{silk98,haehnelt98}. The accretion energy released during the growth of a nuclear black hole vastly exceeds the potential energy of the host and thus must couple inefficiently to the surrounding gas \citep{fabian12,king15}. A wind accelerated and heated by shock fronts whose cooling time scale lies below its expansion timescale conserves momentum \citep{zubovas14a}. In momentum-conserving outflows, the shocked wind radiates away its kinetic energy and compresses into a dense gas. Ram pressure is communicated to the ISM driving outflows to a kiloparsec and perhaps beyond \citep{zubovas12,king15}.  Momentum conserving winds lead to $\sigma^4$ scaling \citep{fabian99,king03,murray05} which lies close to the observed $M-\sigma$ relation \citep{fabian12}.

    If the optical depth of dust in the ISM is greater than that due to Thomson scattering, the coupling between AGN radiation and dust can be significant. In that case, radiation pressure can power an outflow \citep{ishibashi18}. Depending on the optical depth of the medium around the AGN and the dust content, momentum ratios $\dot{p}/(L_{\rm AGN}/c)$ can reach values of up to 10 or higher in sources with high AGN luminosity and high dust content, with large covering fraction leading to higher coupling efficiency. The power ratio ($\dot{E}/L_{\rm AGN}$) can be greater than 0.01.

	During radio mode feedback, jets launched from SMBHs inflate bubbles in the surrounding hot atmosphere, many of which are observed as surface brightness depressions or cavities in X-ray images. Atmospheric heating by a combination of shock fronts \citep{randall11}, turbulence \citep{zhuravleva14}, and enthalpy released as the cavities rise suppresses and often neutralizes cooling and star formation \citep{mcnamara07,mcnamara12,fabian12}.  As the bubbles rise, they lift low entropy, metal-rich gas that is dispersed in the inner several 10s of kiloparsecs of BCG's atmosphere \citep{nulsen02,kirkpatrick15,simionescu08,gitti11}. Some of this gas may cool and condense in the bubbles' updrafts into molecular clouds \citep{revaz08,mcnamara16,voit17}. Molecular gas is plentiful in brightest galaxies (BCGs) residing at the centers of cool-core clusters, groups and galaxies \citep{edge02,salome03,pulido18,russell19,olivares19}. 
	
	In Galactic outflows detected in ionized, neutral, and molecular forms, molecular gas flows usually contain most of the mass.  Molecular clouds directly fuel star formation and nuclear black holes. Therefore, they have the greatest potential impact on the evolution of galaxies.  Several investigations have explored outflow properties including size, mass, velocity, and their relationships to galaxy morphology, AGN luminosity, star formation rate, and stellar mass. \citet{cicone14} found a trend between AGN luminosity and outflow rate. They also found that in starburst galaxies, the ratio of mass outflow rate to star formation rate, dubbed the loading factor, is typically $\sim$ 1--4 but increases to upwards of 100 for the most powerful AGN. 
	
	\citet{fiore17} found that AGN winds are a dominant driving mechanism in massive galaxies at redshifts of two and below.  They concluded tentatively that winds are capable of sweeping or destroying molecular gas in massive galaxies. Most studies have focused on relatively small, heterogeneous samples. \citet{fluetsch18} compiled a sample of relatively nearby galaxies to study these relations with an eye to understanding selection biases. They found that outflows can be both energy-conserving and radiation pressure driven, and that less than five percent of the gas escapes the host galaxies.  No correlation was found between molecular gas outflow rates and radio power. Nevertheless, several instances of jet-powered outflows are known \citep{morganti15,morganti13,mahony16}.

	Radio mode or radio/mechanical feedback is thought to be the agent maintaining the inner cooling regions of hot atmospheres thus preventing star formation in massive galaxies at late times.  ALMA observations of central galaxies in clusters with cooling atmospheres have revealed a far more complex picture \citep{russell19, olivares19}. While stellar populations in most central galaxies are usually old, those in cooling core clusters are instead often rich in molecular gas and star formation \citep{edge01,salome03,odea08,donahue15}. For the past two decades, the Chandra and XMM-Newton observatories have characterized the hot X-ray emitting atmospheres in and around BCGs with fine detail, revealing multi-phase, cooling atmospheres. Detailed observations of the molecular gas content are available for only a dozen systems or so. These studies have revealed molecular gas filaments trailing behind buoyantly rising X-ray cavities and interacting with radio lobes \citep{mcnamara14,russell14,russell16,russell17a,russell17b,vantyghem16,vantyghem18}. These spatial correlations indicate that radio AGN are disrupting and perhaps expelling molecular gas from the central galaxies. Molecular gas may also condense in the updrafts of rising radio bubbles \citep{mcnamara14,mcnamara16,russell19}. 
	
	Here we analyze molecular flows in central galaxies and compare them to those in active galaxies from the \citet{fluetsch18} (Fluetsch). Fluetsch did not include outflows driven by radio-mode feedback in central galaxies.  In fact, most studies neglect radio-mode feedback as an effective mechanism for driving massive outflows.  
	Our goal is to understand the relative impact these disparate modes of feedback can have on the evolution of galaxies.  
	
	In recent years, the Atacama Large Millimeter Array (ALMA) has observed molecular gas in central galaxies with unprecedented spatial and velocity resolution. This combined with sensitive X-ray imaging and nebular line spectroscopy have provided a vivid picture of radio-mode feedback. Such work has revealed some of the most massive molecular gas flows known in the universe.  Here we examine ALMA observations of BCGs to decompose the flow and non-flow components. We investigate the relative efficiency and driving power of different feedback modes and we examine their roles in suppressing star formation. 
	
	Throughout this paper, we adopt $H_{0} =$ 70 km s$^{-1}$ Mpc$^{-1}$, $\Omega_{M} =$ 0.3 and $\Omega_{\lambda} =$ 0.7.
	
	\section{Sample selection}
	\label{sec:sample}
	Archival ALMA observations of 12 BCGs were analyzed. All targets in our sample lie below a redshift of 0.6 and span the range in the molecular gas mass of 10$^7$ -- 10$^{11}$ M$_\odot$. Star formation rate ranges between a few to $\sim$ 600 M$_\odot$ yr$^{-1}$ and radio mechanical AGN power between 10$^{42}$ -- 10$^{46}$ erg s$^{-1}$. Their stellar masses are greater than 10$^{11}$ M$_\odot$. These properties are summarized in table~\ref{tab:bcg_properties}. Most ALMA archival targets were observed based on their large molecular gas reservoirs \citep{edge02,pulido18}. Therefore, our sample is biased, incomplete and does not represent all central cluster galaxies. We consider this bias throughout our analysis. We have included NGC 1275, the central galaxy in the Perseus cluster, observed with the IRAM telescope \citep{salome11}. 
    
    The \cite{fluetsch18} sample of 45 galaxies with gaseous outflows lies within redshifts of $z<0.2$, providing a good reference sample to compare to central galaxies.  They include four ULIRGs for which the molecular properties were determined from far-infrared OH transitions. They classify galaxies as Seyfert, low-ionization nuclear emission-line region (composite) and starburst (star-forming) galaxies based on the BPT-[SII] diagram. They include galaxies with flows in neutral and ionized phases. We refer to these categories as AGN, composite, and starburst galaxies, respectively.
	
	
	\section{Data Analysis}
	We used automated science pipeline scripts provided by ALMA with the data sets to reduce and calibrate the data. For Cycle 0 Early Science data (Abell 1835 and Abell 1664), scripts were modified according to instructions provided on the ALMA website\footnote{\href{https://casaguides.nrao.edu/index.php/Updating_a_script_to_work_with_CASA_4.2}{https://casaguides.nrao.edu/index.php/Updating\_a\_script\_to\_work\_with\_CASA\_4.2}} to re-calibrate the data with \textsc{casa} 4.2. All other data were processed in \textsc{casa} version 4.7.2. Additional flagging was done for some BCGs such as phase center correction (Abell 1835), total flux calibration (Phoenix) and self-calibration was performed on some datasets to improve the signal to noise. 
    
    \textsc{casa} tasks \textsc{uvcontsub} and \textsc{clean} were used to subtract the continuum and generate image data cubes, respectively. We used Briggs weighting with parameter 2 to improve sensitivity to faint filaments. The rms noise in the data cubes was found to be close to the theoretical rms.
	
	
	\section{Comparison of Molecular Flow Sizes, Speeds, Masses and Power}
	The outflows studied by Fluetsch are generally unresolved spatially due to their small angular sizes.  Their masses and velocities were measured by fitting at least two gaussians to the molecular gas velocity profiles. The area and width of the broader gaussian component provided mass and velocity estimates.  In some cases, the flows are resolved and visibly surround the galaxy.  However, their morphologies are usually indiscernible due to their small angular sizes. 
	
	In contrast, the flows in BCGs are well resolved with a variety of morphologies and velocity patterns.   
	Molecular gas in BCGs is often filamentary, extending out from the center with multiple spectral components \citep{russell19,olivares19}. In most BCGs, the molecular gas lies outside of the nucleus in an unsettled state. Molecular gas is rarely observed in ordered motion about the nucleus such as disks. We consider this off-nucleus gas as a flow in BCGs. Some of this gas may be flowing towards the BCG as opposed to a pure outflow. A simple, comprehensive characterization of their flows that can be compared to the \citet{fluetsch18} systems would be desirable.
	
	Although some filaments have smooth velocity gradients, they are much more extended compared to the sizes of molecular gas disks observed in galaxies \citep{rose19a, boizelle17}. Their position-velocity (PV) diagrams do not show the characteristic `S'-shaped curve that represents rotation (see Fig.~\ref{pv}). Some of the molecular gas in A1664 may be forming a disk of molecular gas in the centre \citep{russell14}. Similarly, the circumnuclear gas reservoir in phoenix has a smooth velocity gradient from $-$200 to 200 km s$^{-1}$ suggestive of a disk \citep{russell17a}. But the extended filaments that we consider as flows have distinct velocity structures from the circumnuclear gas. Molecular gas disks are detected in very few BCGs such as HydraA \citep{rose19a} and Abell 262 \citep{russell19}. We have excluded those BCGs from our sample.
	In the following subsections, we describe the process of estimating flow properties in BCGs and compare them to flows in the Fluetsch sample.
	
	\subsection{Flow velocities}
	\label{speed}
	
    Spectra from data cubes were extracted in beam-sized regions centered on each pixel in the entire emission region. They were fitted with a model consisting of one Gaussian component using the \textsc{lmfit} Python package\footnote{\href{https://lmfit.github.io/lmfit-py/}{https://lmfit.github.io/lmfit-py/}}. A significance of greater than $3\sigma$ was imposed for the detection of line emission, evaluated by performing 1000 Monte Carlo iterations. Integrated CO images and velocity centroid maps of BCGs analyzed here are presented in Appendix 1.
    
    \cite{fluetsch18} identified outflows spectroscopically, based primarily on the detection of two velocity components in the spectrum of the entire emission region. They calculated flow speed using FWHM$_{\rm broad}/2 + |v_{\rm broad} - v_{\rm narrow}|$, where FWHM$_{\rm broad}$ is the full width at half maximum of the broad component in the extracted spectrum of the source and $v_{\rm broad}$ and $v_{\rm narrow}$ are the mean speeds of broad and narrow components. This method can be applied only when narrow and broad components are identified. In the absence of multiple components, Fluetsch concluded non-detection of an outflow. To allow for a comparison to their method, we extracted spectra in the entire emission region including all emission structures and fit them with one or two Gaussian components. The prescription described above was applied to estimate outflow velocities in BCGs according to their method.
    
    The outflows in many Fluetsch galaxies are barely resolved spectrally and few are resolved spatially.  Therefore, the outflow speeds and gas masses must be large enough for the spectrum to deviate from gaussianity. Applying this method to Abell 2597, Abell 1835, AS1101 and NGC4696, the summed spectra deviate only marginally from gaussianity.  Their flows would therefore have been missed were they too distant to be spatially resolved. Nevertheless, some of these systems harbor some of the most massive molecular gas flows known.
	
\begin{figure}
		\centering
		\includegraphics[width=0.4\textwidth]{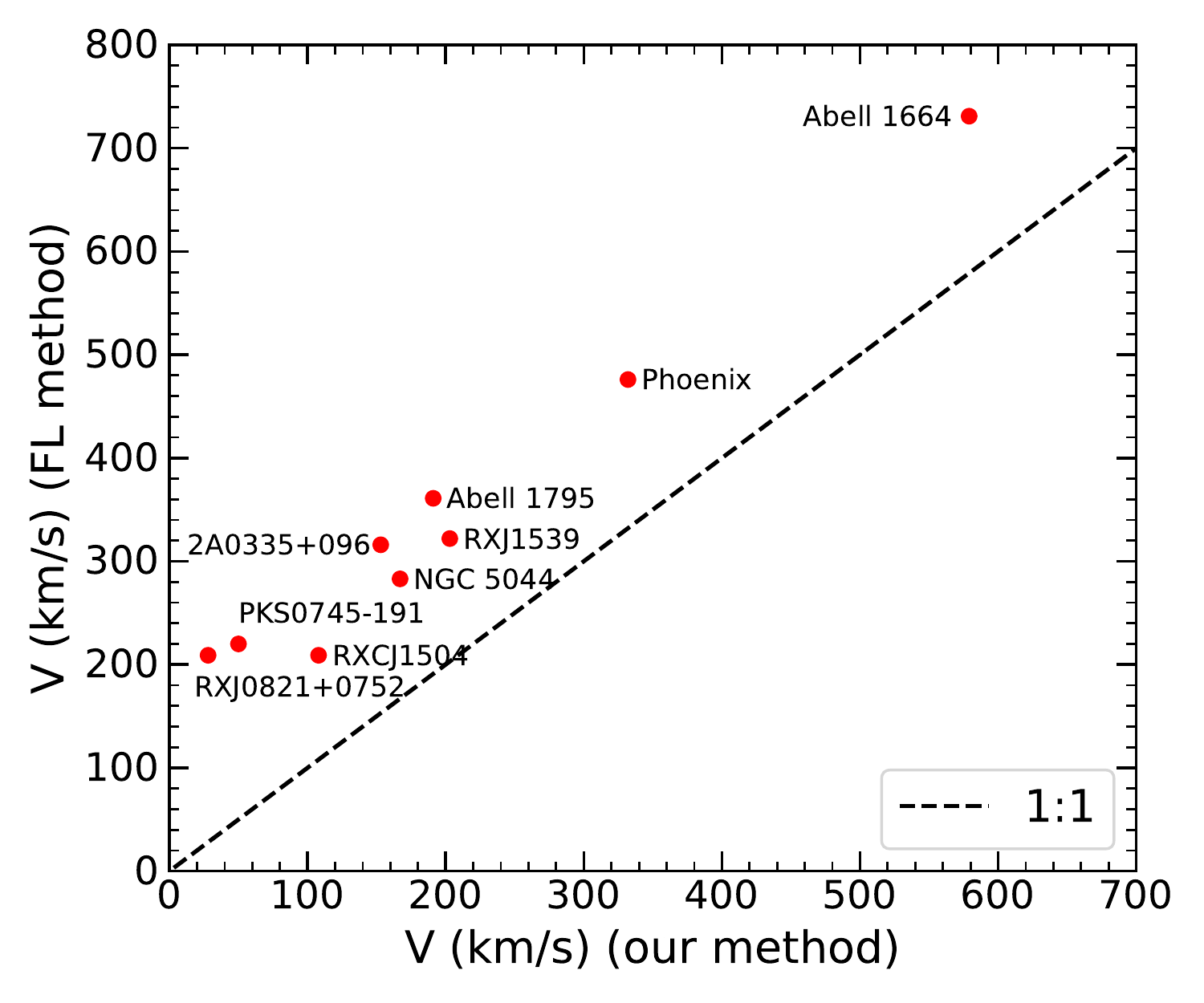}
		\caption{Comparison of flow speeds estimated using the Fluetsch method against our method. The black dashed line is the 1:1 relation.}
		\label{v_comp}
	\end{figure}

	We identify gas flows in BCGs and determine their velocities and sizes individually from spatially extended emission components. Gas speeds were estimated as flux weighted average speed along the spatially extended filaments or clumps of gas from their velocity maps within regions we considered as flows. The regions and the corresponding velocity maps are shown in the Appendix. If more than one filament or clump is present, the average speeds of all extended components are considered as the flow speed. Figure~\ref{v_comp} compares Fluetsch's methodology to ours in systems where both methods can be applied. The flow speeds estimated using the two methods are correlated with a Pearson correlation coefficient of 0.77 and a p-value of 0.001. Speeds estimated using the Fluetsch method are two times higher compared to our method, except for RXCJ0821. The molecular gas in RXCJ0821 lies in two clumps north of the BCG's nucleus. Their mean velocities are close to the BCG's systemic velocity. However, their FWHMs are significantly larger than their mean relative velocities, yielding much higher flow velocities using the Fleutsch method. The Fluetsch method captures a small fraction of the gas that is flowing at high speeds and overestimates the flow velocity. In what follows, we adopt speeds estimated using our method. 
	However, were we to adjust Fluetsch velocities to velocities obtained using our method by dividing them by a factor of two, it does not change our results qualitatively (see Appendix~\ref{appB}).
	
\begin{figure}
		\centering
		\includegraphics[width=0.4\textwidth]{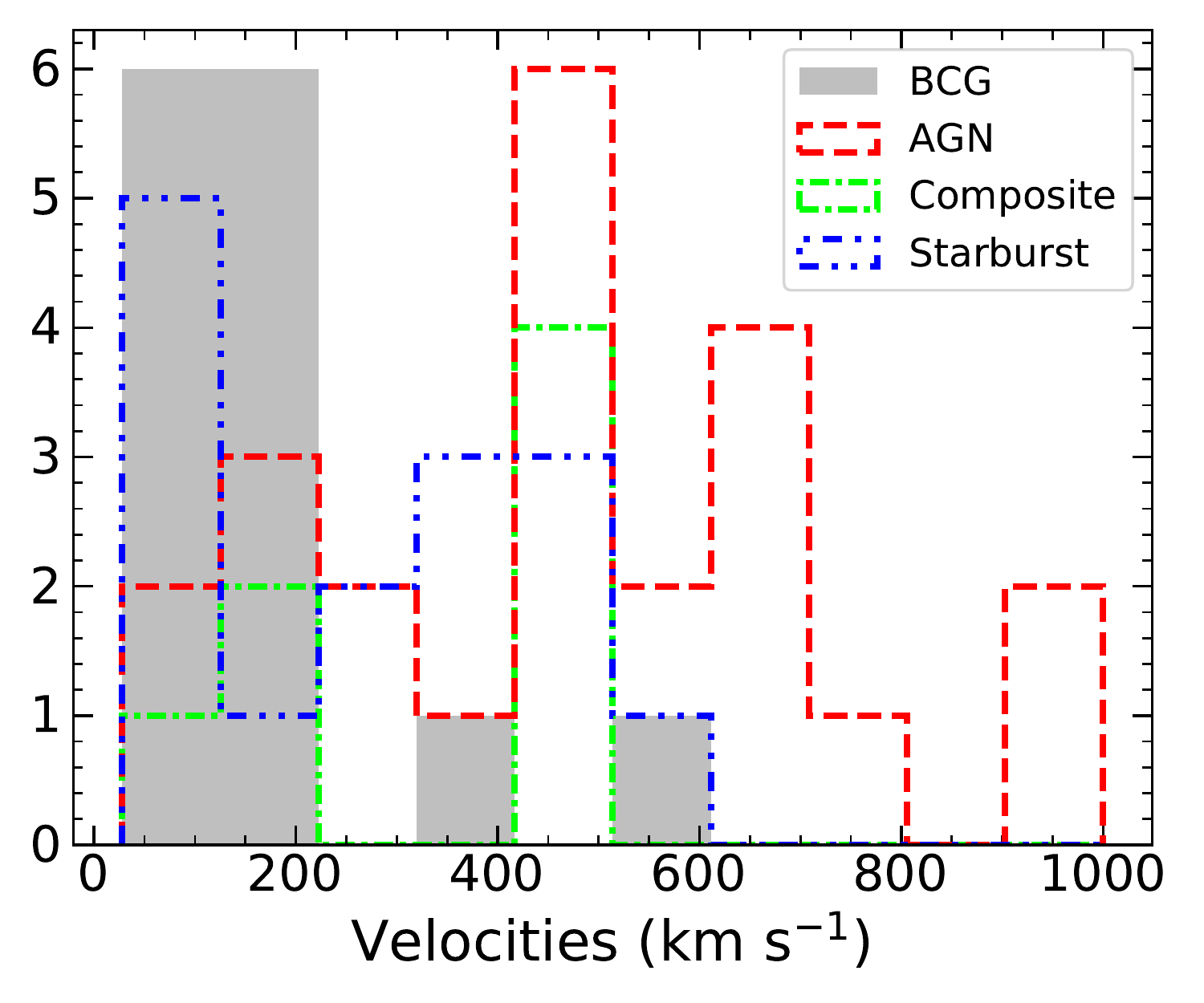}
		\caption{The distribution of flow speeds in BCG, AGN hosting, composite and starburst galaxies. The flows in BCGs have relatively lower speeds compared to AGN hosting galaxies.}
		\label{velhist}
	\end{figure}

	Figure~\ref{velhist} shows the distribution of flow speeds for all systems.  BCG flow speeds lie in the range 50--300 km s$^{-1}$ with a mean of 173 km s$^{-1}$. Abell 1664 with a speed of 579 km s$^{-1}$ is a moderate outlier. By comparison, the flow speeds of starburst and AGN galaxies lie in the range 50--600 km s$^{-1}$ and 100--800 km s$^{-1}$, respectively. The mean outflow velocity for AGN and starburst galaxies is 447 km s$^{-1}$ and 243 km s$^{-1}$, respectively. Thus BCG flow speeds on average lie $\sim$60\% below AGN hosting galaxies but are similar to starburst galaxies.  
	
	\subsection{Flow sizes}
	\label{size}
	
	Flow projected sizes are determined from the projected lengths of the filaments in ALMA CO images. 
	The velocity profiles of the gas in filaments in most BCGs are distinct from the gas in the central regions of BCGs. Filamentary gas in cluster central galaxies generally has narrow velocity widths with full width at half maximum velocities  $\lesssim $100km s$^{-1}$ and mean recessional speeds of several hundred km s$^{-1}$. These filaments are considered off-nuclear flows.  To examine the filaments we imaged the data cube over the velocity ranges observed in the filament. We then measured the filament sizes from the BCG nucleus to the most distant part of the filament. When multiple filaments are present their sizes are averaged. For example, PKS0745 has three long filaments in the SE, NW and SW directions. We take the length of each filament from the BCG center marked as x in Figure~\ref{maps} to the average outermost edge of the filament as the size of the flow.
	
\begin{figure}
		\centering
		\includegraphics[width=0.4\textwidth]{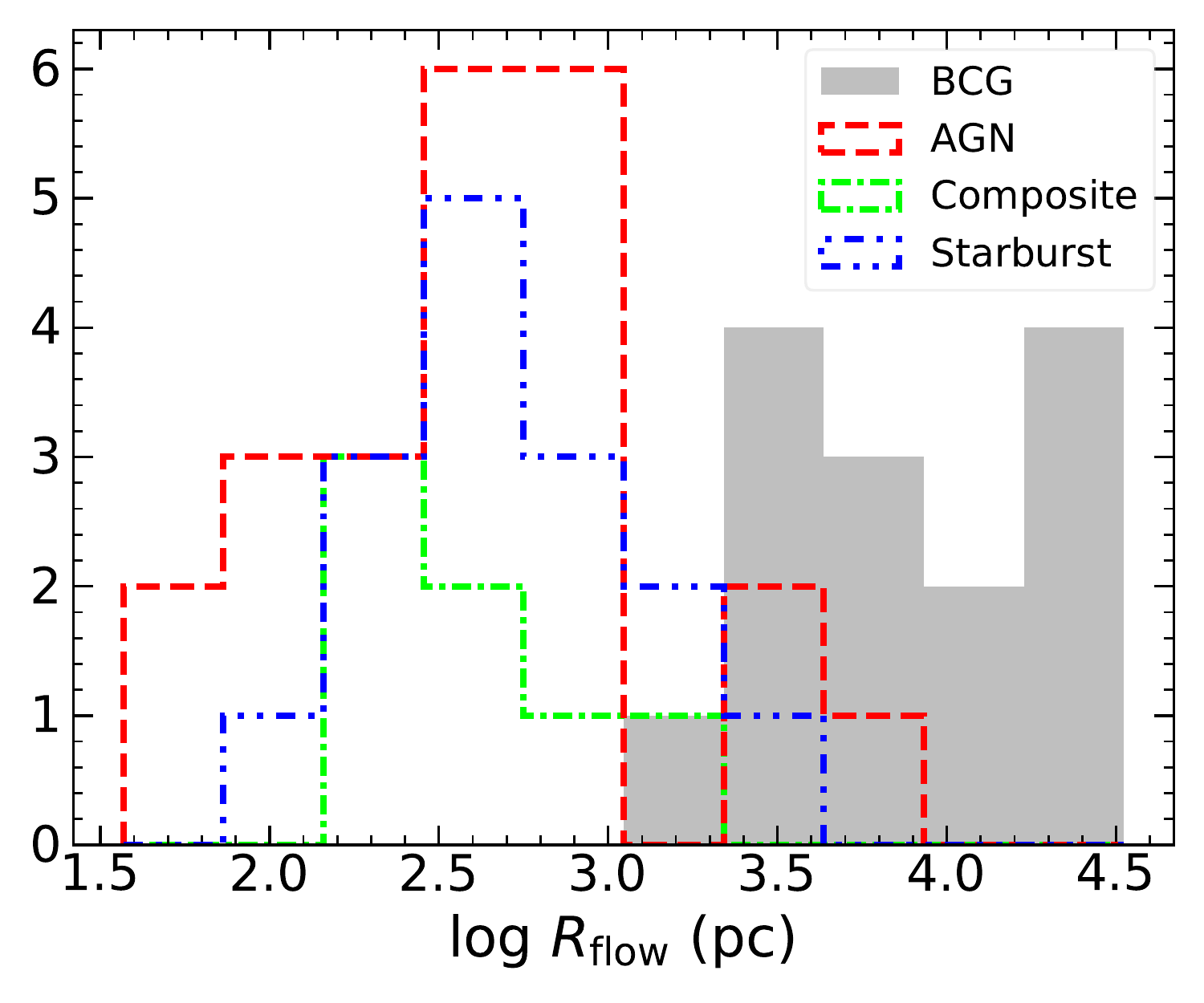}
		\caption{Histogram of flow radii in BCGs, AGN, composites and starburst galaxies.}
		\label{sizehist}
	\end{figure}

	In Figure~\ref{sizehist} a histogram of flow size in BCGs, AGN hosting, composites and starburst galaxies are presented. BCG flow sizes generally range between 2 and 15 kpc. Phoenix, RXCJ1539 and NGC1275 are exceptions with flows extending out to $\sim$21, 24 and 33 kpc, respectively. The average molecular flow size is $\sim$12 kpc in BCGs. AGNs span a broad range of flow sizes from 30 pc to 3 kpc.  They are generally much smaller and more compact than those in BCGs with an average flow size of only $\sim$1 kpc. Molecular flow sizes in starburst and composite galaxies range between 100 pc and 1 kpc. Thus, the molecular flows in active galaxies studied by Fluetsch are generally confined to the nuclear regions.  In contrast, the molecular flows in BCGs tend to be 10 to 1000 times larger extending well into the bulge and beyond.

	\subsection{Flow masses}
	\label{mass}
	Molecular gas masses in BCGs are calculated using the integrated intensity of the CO line emission. Flow masses were estimated from the CO intensity of spectra extracted from flow regions. The total molecular gas mass was estimated using the spectrum extracted from a region enclosing all detected CO emission. The molecular gas mass is inferred using an empirical expression calibrated for the CO(1-0) line. The integrated line flux ratios of $I_{\rm CO(2-1)}/I_{\rm CO(1-0)}$ $\approx$ 3.2 and $I_{\rm CO(3-2)}/I_{\rm CO(1-0)}$ $\approx$ 7.2 are used to convert integrated CO(2-1) and CO(3-2) flux densities to CO(1-0), respectively. The corresponding brightness temperature ratios are CO(2-1)/CO(1-0) = 0.8 and CO(3-2)/CO(1-0) = 0.8. These ratios are based on observed ratios at CO(3-2), CO(1-0) or CO(2-1) and CO(1-0) in BCGs \citep{russell16,vantyghem16,vantyghem17,vantyghem18}. They correspond to an excitation temperature of $\sim$20--25 K and high densities $\sim$ 10${^4}$ cm$^{-3}$. Integrated CO(1-0) flux is converted into molecular gas mass using the equation \citep{solomon05,bolatto13}
	\begin{equation}
	\begin{split}
	M_{\rm mol} = & 1.05 \times 10^{4} \Bigg(\frac{X_{\rm CO}}{X_{\rm CO,Gal}}\Bigg) \bigg(\frac{1}{1+z}\bigg)\\
	& \bigg(\frac{S_{\rm CO} \Delta v}{\rm Jy\, km\, s^{-1}}\bigg) \bigg(\frac{D_{\rm L}}{\rm Mpc}\bigg)^{2}\, {\rm M_{\odot}},
	\end{split}
	\end{equation}
	where $S_{\rm CO} \Delta v$ is the integrated flux density of the CO(1-0) line, $D_{\rm L}$ is the luminosity distance, $z$ is the redshift of the galaxy and $X_{\rm CO}$ is the CO-to-H$_2$ conversion factor, with $X_{\rm CO,Gal}$ = 2$\times$10$^{20}$ cm$^{-2}$ (K km s$^{-1}$)$^{-1}$.
	
	The adopted value of $X_{\rm CO}$ lends to a factor of two or more uncertainty in the gas mass estimates. Its value depends on temperature, density and particularly, the metallicity of the molecular gas \citep{bolatto13}. Estimates of $X_{\rm CO}$ are available for Milky Way ($X_{\rm CO,Gal}$) and nearby galaxies. No direct estimate of $X_{\rm CO}$ is available for BCGs. However, studies have shown that the hot atmospheres surrounding BCGs, from which the molecular clouds have likely condensed,  have metallicities lying between $\sim$0.6--0.8 $Z_{\odot}$, close to if not below solar metallicity. Furthermore, the line widths of individual molecular gas clouds are comparable to those in the Milky Way \citep{rose19,tremblay16, david14, heyer15}. CO line ratios are consistent with optically thick molecular gas ~\citep{russell19}. \citet{vantyghem17} found that the optically thin $^{13}$CO to optically thick $^{12}$CO line ratio in one BCG indicated abundances lying within a factor of two of the Galactic value.  
	Therefore, we have adopted $X_{\rm CO,Gal}$ for BCGs in our calculations apart from Phoenix.
	
	The BCG in the Phoenix cluster is undergoing star formation at a rate of $\sim$500 -- 800 M$_\odot$ yr$^{-1}$ \citep{mcdonald12a} similar to LIRG/ULIRGs. It has been shown that the use of $X_{\rm CO,Gal}$ in ULIRGs may overestimate the amount of molecular gas by a factor of five~\citep{downes93,solomon97,downes98}. Therefore, we have adopted $X_{\rm CO}$ = 0.4$\times$10$^{20}$ cm$^{-2}$ (K km s$^{-1}$)$^{-1}$ for Phoenix following \citet{russell17a}. \citet{fluetsch18} use $X_{\rm CO,Gal}$ in their calculations for all galaxies except for LIRGs and ULIRGs for which they use $X_{\rm CO}$ of 0.4$\times$10$^{20}$ cm$^{-2}$ (K km s$^{-1}$)$^{-1}$ and 0.6$\times$10$^{20}$ cm$^{-2}$ (K km s$^{-1}$)$^{-1}$, respectively, which is in agreement with the factor we assume.
	
	Radio interferometers such as ALMA can underestimate the total flux if the angular size of the emission region is larger than the maximum recoverable scale given by $\sim$0.6$\lambda$/D$_{\rm min}$, where $\lambda$ is the wavelength of the observation and D$_{\rm min}$ is the shortest baseline. Single dish telescopes generally do not suffer from this problem due to their large maximum recoverable scales. Therefore, we compared our molecular gas mass measurements with molecular gas masses obtained from single-dish observations \citep{edge01} using the same CO-to-H$_2$ conversion factor as ours to quantify the missing flux in our observations.  The ALMA  molecular gas masses are within a factor of the single-dish molecular gas mass measurements. ALMA detected 60 percent more molecular gas in RXCJ0821, while in A1664, A1835 and NGC5044 the ALMA masses lie 25 to 40 percent below the single dish mass measurements. In the latter systems, ALMA may have resolved away some extended molecular gas. These discrepancies do not qualitatively affect our results.
	
\begin{figure}
		\centering
		\includegraphics[width=0.4\textwidth]{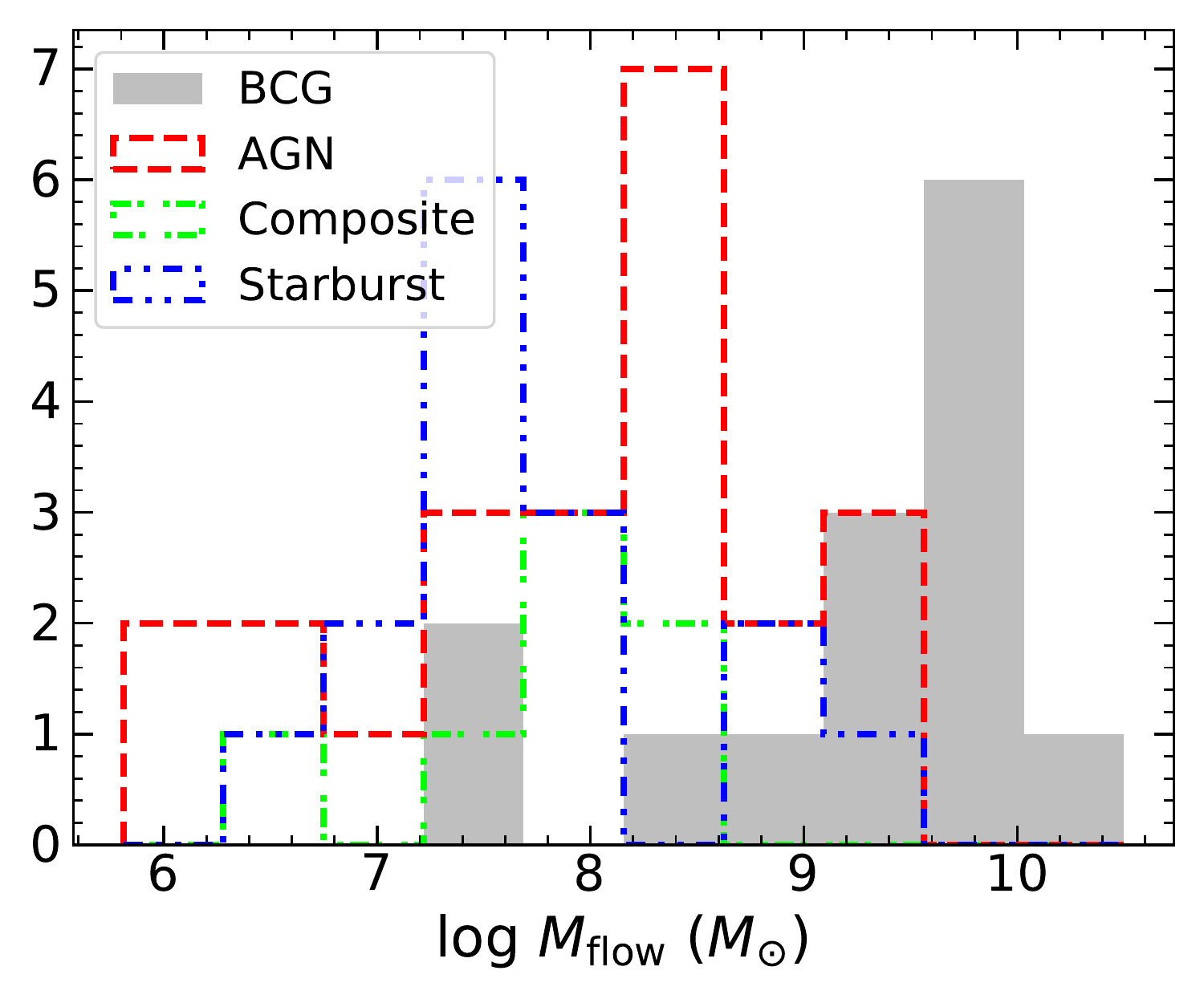}
		\caption{Histogram of the mass of molecular gas in outflows of BCGs, AGNs, composites and starburst galaxies.}
		\label{masshist}
	\end{figure}

	Figure~\ref{masshist} shows the molecular gas mass associated with the out(in)flows on a logarithmic scale. BCG molecular flow masses lie between 0.3--12 $\times 10^{9}$ M$_{\odot}$ apart from the much smaller masses in NGC 5044 and NGC4696 of 10$^{7.7}$ M$_{\odot}$ and 10$^{7.5}$ M$_{\odot}$, respectively. BCG flow masses differ markedly from those in \citet{fluetsch18}. BCGs possess on average $\sim$100 times more molecular gas in flows compared to composites (10$^{7.5}$--10$^{8.5}$ M$_{\odot}$) and starburst galaxies (10$^{6}$--10$^{8}$ M$_{\odot}$). On the other hand, a broad mass range is found in AGN galaxies, lying between $10^{6}$--$10^{9}$ M$_{\odot}$. Nevertheless, the most massive flows in BCGs exceed those in AGN by more than a factor of 10.
	
\begin{figure}
		\centering
		\includegraphics[width=0.4\textwidth]{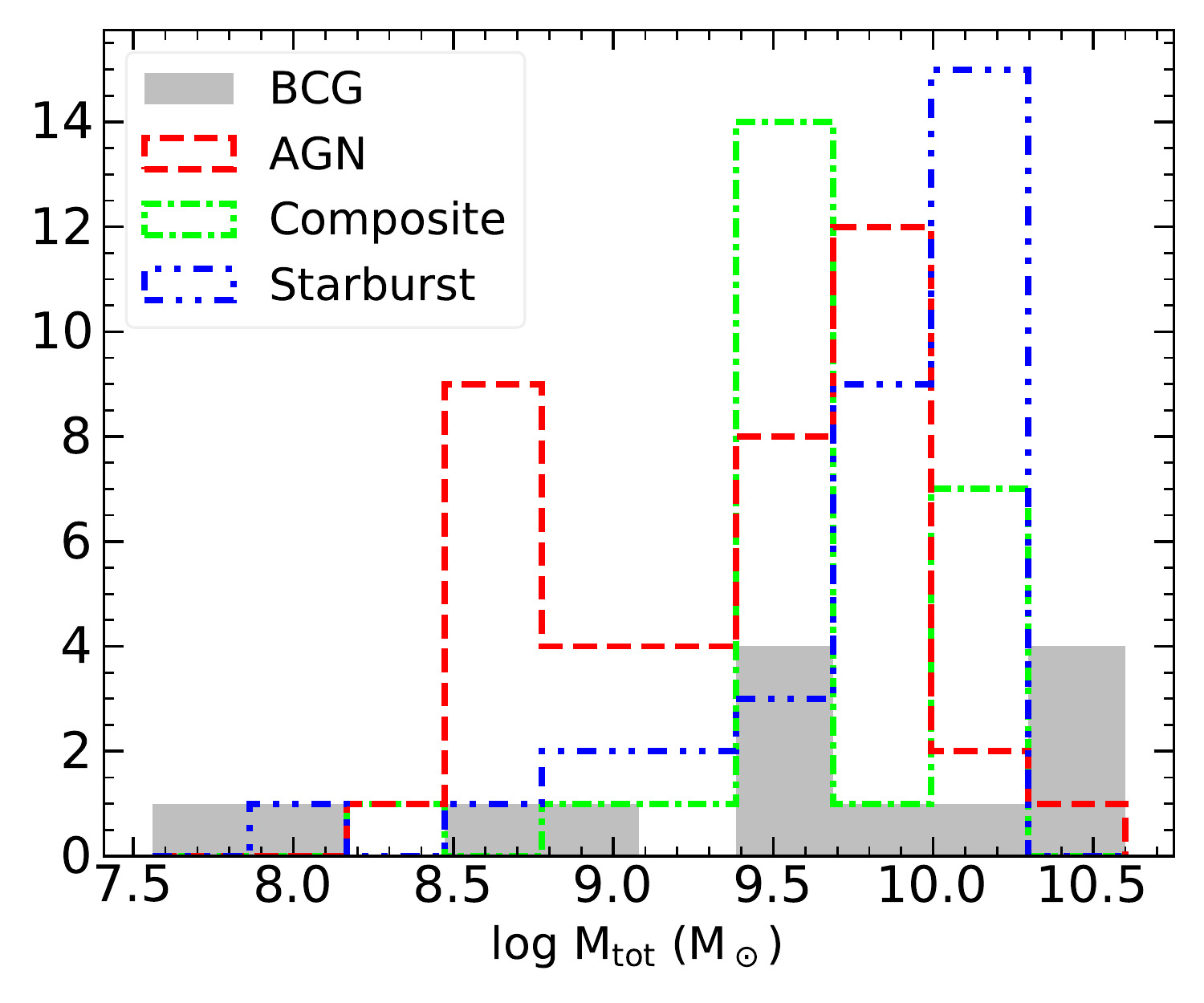}
		\caption{Histogram of total molecular gas mass in BCGs, AGNs, composites and starburst galaxies.}
		\label{mtothist}
	\end{figure}

	However, the total molecular gas masses in BCGs are comparable to those in 
	\citet{fluetsch18} (Fig~\ref{mtothist}). By comparison, the average total molecular gas mass in BCGs is $\sim$10$^{10}$ M$_\odot$, while it is 5$\times 10^9$ M$_\odot$ in AGN and 3$\times 10^9$ M$_\odot$ in composite and starburst galaxies. 
	
	About half of the total molecular gas mass in BCGs is found in filaments. Some BCGs like Perseus are filament-dominated, in which almost all of the molecular gas is in filaments. In AGN hosting galaxies the amount of molecular gas in the outflow is only $\sim$10 percent. It is even lower in composite and starburst galaxies at $\sim$3 percent.
	
	On a cautionary note, many BCGs analyzed here were selected for observation based on their high molecular gas masses during the early ALMA cycles. Therefore they do not represent the molecular gas masses of average BCGs.  We have kept this in mind as we draw scientific inferences from the data.

	\subsection{Measurement uncertainties}
	The true sizes of molecular flows and their velocities are systematically underestimated due to projection.  Furthermore, when observed in emission we do not know what side of the nucleus the filaments are located. Unless absorption against the continuum is observed, it is very difficult to determine whether the gas is moving towards or away from the AGN \citep{rose19}. Therefore, whether the filaments in an individual object are flowing in or out is unknown. Both inflows and outflows can be present at the same time. Observed velocities will be lower if both inflowing and outflowing gas is present along the line of sight. For example, in PKS0745, flux weighted average velocity is only 50 km s$^{-1}$, however, there is a component at $\sim$200 km s$^{-1}$ in the extended filament. The low-velocity gas could have detached from the flow and slowed down or infalling onto the central galaxy. These effects would introduce a factor of a few uncertainty in velocity in most objects. Although these effects are difficult to quantify, we adopt 0.3 dex uncertainty in velocity.
	
	The primary mass uncertainty comes from the application of $X_{\rm CO}$, which contributes 0.3 dex \citep{bolatto13}. Statistical uncertainties due to the fitting the total intensities add about $\lesssim$20 percent. Thus the average error on the mass of the outflow may be as large as 0.31 dex. Orientation and projection introduce scatter in both velocity and radius measurements. Including this scatter, the overall average error on flow properties is $\sim$0.5 dex.

	\begin{table*}
		\centering
		\small
		\label{tab:bcg_properties}
		\begin{tabular}{ c c c c c c c c c c c c c c} 
			\hline
			BCG & $z$ & log $M_{\rm flow}$ & log $M_{\rm tot}$ & $R_{\rm flow}$ & $v_{\rm flow}$ & $\dot{M}_{\rm flow}$ & log $P_{\rm Mech}$ & log $L_{\rm Nuc}$ & SFR & log ($M_{\star}$) & Ref.\\
			 & & ($M_{\odot}$) & ($M_{\odot}$) & (kpc) & (km s$^{-1}$) & ($M_{\odot}\,{\rm yr^{-1}}$) & (erg s$^{-1}$) & (erg s$^{-1}$) & ($M_{\odot}\,{\rm yr^{-1}}$) & ($M_{\odot}$) & \\
			(1) & (2) & (3) & (4) & (5) & (6) & (7) & (8) & (9) & (10) & (11) & (12) \\
			\hline
			2A0335+096 & 0.0346 & 8.31 & 8.59 & 2 & 153 & 17 & 43.949 & $<40.80$ & 0.47 & 11.73 & [a] \\
			Abell 1664 & 0.128 & 9.56 & 9.87 & 6.3 & 579 & 341 & 44.04 & $<41.50$ & 13.18 & 11.1 & [b] \\
			Abell 1795 & 0.063 & 9.36 & 9.52 & 8.3 & 191 & 54 & 43.133 & $<41.24$ & 3.47 & 11.84 & [c] \\
			Abell 1835 & 0.252 & 10 & 10.6 & 12 & 61 & 52 & 45 & $<42.78$ & 117.49 & 10.57 & [d] \\
			Abell 2597 & 0.0821 & 10 & 10.3 & 15 & 61 & 42 & 44.278 & $<41.73$ & 3.98 & 11.51 & [e] \\
			PKS0745-191 & 0.1028 & 9.7 & 9.89 & 4.1 & 50 & 63 & 45.7 & 42.11 & 13.49 & 11.71 & [f] \\
			Phoenix & 0.596 & 10.5 & 11.1 & 21.4 & 332 & 501 & 45.85 & 44.37 & 616.6 & 12.48 & [g] \\
			RXCJ1504 & 0.216902 & 9.7 & 10.4 & 18 & 108 & 30 & 44.96 & 42.66 & 85.11 & 11.7 & [h] \\
			RXCJ0821 & 0.109 & 10.21 & 10.32 & 4 & 28 & 72 & 44.4 & $<41.14$ & 37 & 11.06 & [i] \\
			NGC5044 & 0.00928 & 7.4 & 7.56 & 2.8 & 167 & 2 & 42.78 & 39.70 & 0.22 & 11.38 & [j] \\
			NGC4696 & 0.00987 & 7.52 & 7.95 & 4 & 222 & 2 & 43.11 & $<39.67$ & 0.16 & 11.79 & [k]\\
            RXCJ1539 & 0.075766 & 9.51 & 10.11 & 23 & 203 & 29 & 42.76 &  & 1.86 & 11.36 & [k] \\
            Abell S1101 & 0.05639 & 9.03 & 9.06 & 8.3 & 160 & 21 & 44.89 & $<40.99$ & 1 & & [k] \\
			NGC1275 & 0.01756 & 9.02 & 9.68 & 33.3 & 117 & 17 & 43.9 & 42.71 & 70.79 & 11.38 & [l,m] \\
            \hline
		\end{tabular}
		\caption{Properties of BCGs in the sample. \textbf{Columns:} (1) BCG name, (2) redshift, (3) mass of the flow, (4) total molecular gas mass, (5) radius of the flow, (6) speed of the flow, (7) flow rate, (8) mechanical power of the AGN, (9) nuclear 2-10 keV luminosity, (10) star formation rate, (11) stellar masses estimated from 2MASS K-band mangitudes following \citet{main17}, (12) references: [a] \citet{vantyghem16}; [b] \citet{russell14}; [c] \citet{russell17b}; [d] \citet{mcnamara14}; [e] \citet{tremblay16}; [f] \citet{russell16}; [g] \citet{russell17a}; [h] \citet{vantyghem18}; [i] \citet{vantyghem17}; [j] \citet{david14}; [k] \citet{olivares19}; [l] \citet{salome11}; [m] \citet{lim08}.}
		
	\end{table*}
	

	\section{Energy, Momentum, AGN radiation}
	\label{sec:energetics}
	AGN release mechanical energy and radiation capable of displacing the gas around them. In ULIRGs and quasars, AGN power is correlated with the kinetic power of the flows \citep{cicone14}, consistent with the AGN's energetic output driving the flows.  
	The output is in the form of fast nuclear winds and radiation, and radiation pressure from young stars. 
	In this section, we compare molecular flow energy and momentum fluxes in BCGs to the radiative energy and momentum fluxes of their AGN.  We then compare those relations with AGN, composite and starburst galaxies to study the difference between these systems.
    
    \subsection{Flow momentum flux vs $L_{\rm AGN}/c$} 
    
    Apart from a few known quasars hosted in BCGs with radio bubbles, the mechanical AGN power dominates the power budget usually by an order of magnitude or more \citep{russell13}. Nevertheless, some systems with powerful radio activity also emit nuclear X-rays.  For the sake of completeness, we compare the nuclear radiation emerging from BCGs to that of the active galaxies in Fleutsch.  
    
    Here, $L_{\rm AGN}$ denotes the total isotropic radiated power emerging from the BCG's nucleus. \citet{russell13} estimated the nuclear X-ray luminosities ($L_{\rm nuc}$) in the $2-10$ keV energy range using photometric and spectroscopic methods. For those systems, we adopted their photometric value for $L_{\rm nuc}$. For the remaining sources (Phoenix and RXCJ0821), we used archival $Chandra$ observations to estimate nuclear luminosity/upper limits using Russell's photometric method. The presence or absence of a nuclear point source was verified by generating an image in the 3--7 keV band and visually inspecting for a central point source.  When absent, upper limits for nuclear luminosities are presented. $L_{\rm nuc}$ was converted to total bolometric AGN radiative luminosity ($L_{\rm AGN}$) by multiplying it by a bolometric correction. For Compton thick AGN, the bolometric correction is $\kappa_{bol} \sim 30$, with an intrinsic uncertainty of 0.2 dex \citep{brightman17}. Our targets are likely Compton thin, so we expect the bolometric correction to be smaller than the one for Compton thick AGN.  Low luminosity type-2 AGN have X-ray bolometric correction factors of $\sim$10 \citep{lusso12}.  Therefore, we adopt $\kappa = 10$ for our calculations. AGN X-ray luminosity is sensitive to short-term AGN variability and can be underestimated.
	
    \label{momentumrate}
    The flow momentum flux is calculated as $\dot{M} v$, where $\dot{M}$ is the mass flow rate in gm s$^{-1}$ and $v$ is the flow speed in cm s$^{-1}$. $\dot{M}$ is estimated by dividing the mass of the flow by the time it takes the flow to reach its projected size ($R/v$). The rate of momentum output in radiation from the AGN is $L_{\rm AGN}/c$.
    
\begin{figure}
		\centering
		\includegraphics[width=0.4\textwidth]{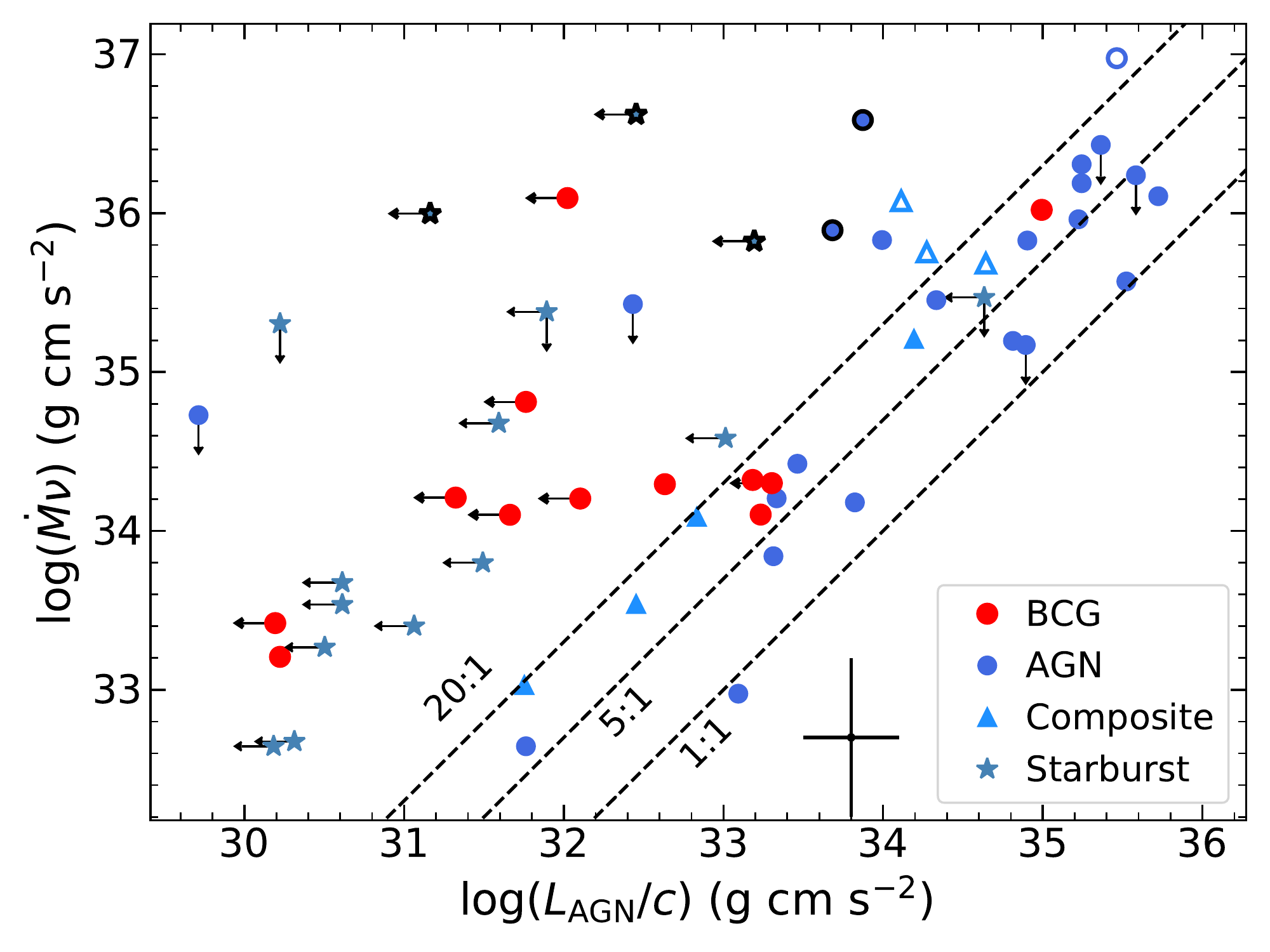}
		\caption{Relationship between the flow momentum rate and the AGN radiation momentum rate. The three dashed lines represent nuclear wind-driven model predictions for energy conserving (20:1), momentum conserving (1:1), and radiation pressure-driven (5:1) flows, respectively. Red circles are denoted by BCGs, AGNs by blue circles, composites by light blue triangles and star-forming galaxies by blue stars. Symbols with white circles in the middle are galaxies with flows detected in OH and symbols with black marker edges are fossil flows from the Fluetsch sample, respectively. The black point with error bars in the bottom right corner represents the average error bar on each point.}
		\label{forces}
	\end{figure}

    In figure~\ref{forces}, the flow momentum flux is plotted against radiative momentum flux $L_{\rm AGN}/c$. The diagram is intended to probe the ability of radiation emitted by the nucleus to drive a flow. Indicated in the figure are flow to AGN radiation momentum flux ratios $\dot{M} v /(L_{\rm AGN}/c)=1,5,20$. According to nuclear wind-driven models, systems in which the ratio lies between 1 and 5 are able to be driven by radiation. Higher values of momentum flux ratios can be obtained by radiation driven flows when the gas in the central regions has high IR optical depth, as in highly obscured AGNs and ULIRGs. However, at kpc scales, the optical depth of the medium is generally too low for outflows to attain momentum flux ratios greater than 5. The energy conserving flows generally have momentum flux ratios above 5. All points lie above the one-to-one line, indicating that radiation is generally unable to drive the flows in these systems. These quantities are correlated in AGN and composite galaxies with a Pearson-r correlation coefficient of 0.73 at greater than 99 percent significance, including systems with either $L_{\rm AGN}$ or $\dot{M}$ upper limits. This indicates that even if AGN radiation pressure is unable to power a flow on its own, it may contribute significantly in high power AGN \citep{fluetsch18}.

    As expected, BCGs show no correlation between the flow momentum flux and the AGN radiative momentum flux (Pearson r-value of 0.24, the 50 percent confidence level). This lack of correlation suggests that radiation pressure has little influence on flows in BCGs.

    The momentum ratios for AGN and composite galaxies lie in the range predicted by nuclear wind-driven outflow models discussed in section~\ref{models_discus}. Still, the large scatter in the relationship makes it difficult to determine whether flows in these systems are energy conserving, momentum conserving, or a combination of both. In BCGs, the rate of change of flow momentum exceeds the radiative momentum input by $\ge 20$, more than the maximum theoretical prediction for energy conserving flows. The only exception is Phoenix, which has a ratio of $\sim$0.2. Phoenix is the most powerful and highly star-forming BCG. It has an active AGN depositing a large amount of energy into its surrounding medium in both mechanical and radiative form in nearly equal amounts. That is reflected in the plot above, where the nuclear luminosity of Phoenix is $\sim 10^{46}$ erg s$^{-1}$, similar to quasars. Phoenix also has a high star formation rate \citep{mcdonald19}. Therefore, the conditions in the Phoenix BCG are similar to those in ULIRGs and quasars. Therefore, Phoenix is expected to be like ULIRGs, with a radiation pressure driven, momentum conserving flow. However, the molecular gas in Phoenix is closely tied to its radio bubbles, rather than star formation or the quasar. The bubbles are apparently doing most if not all of the work.
	
	\subsection{Momentum}
	\label{mom}
\begin{figure}
		\centering
		\includegraphics[width=0.4\textwidth]{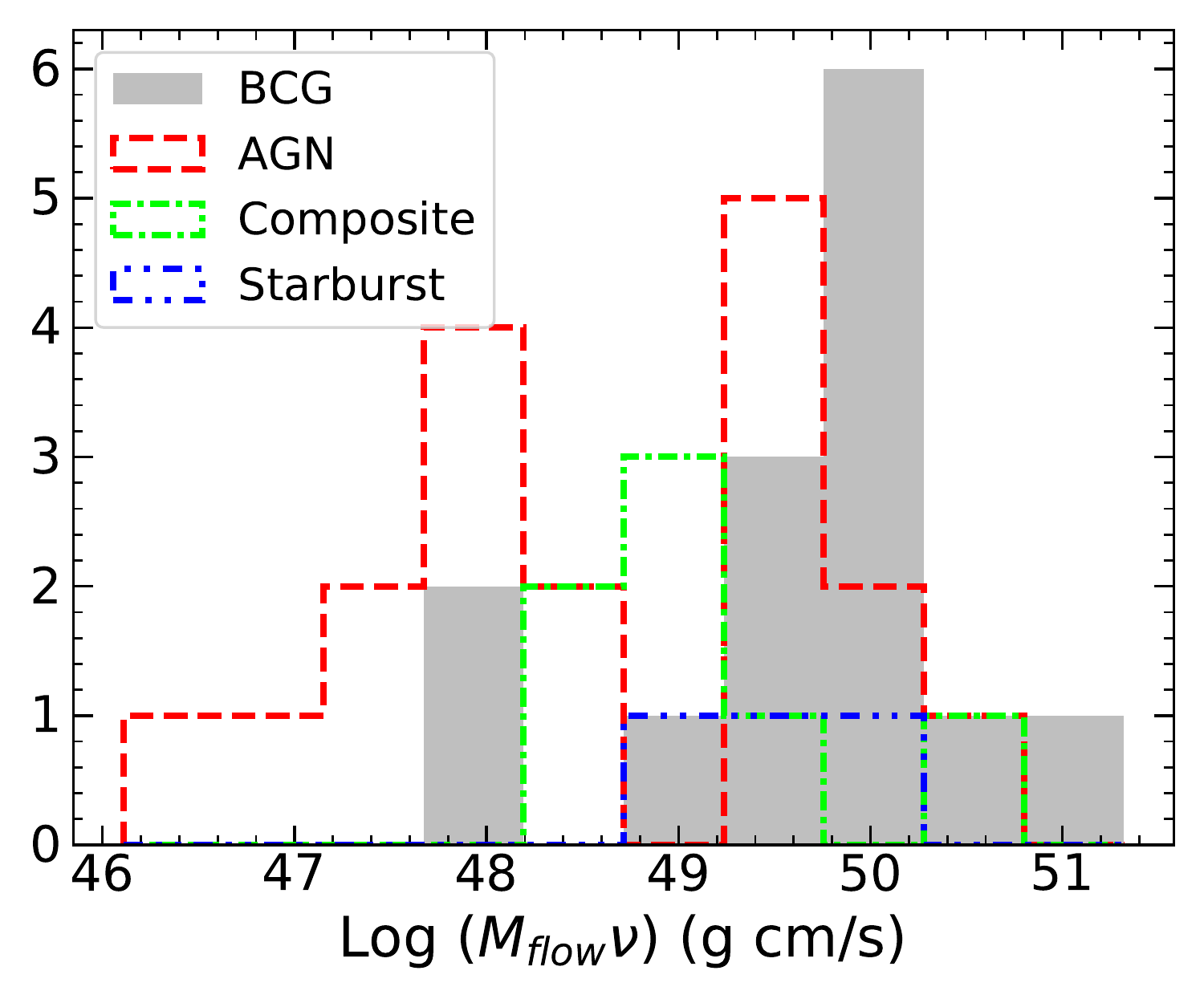}
		\caption{The histogram of flow momenta ($M_{\rm flow} \left\langle v_{\rm flow}\right\rangle $) in BCGs, AGN, composite and starburst galaxies.}
		\label{mom-hist}
	\end{figure}

	Momentum may be conserved in some gas flows. After the hot shocked gas radiates away most of the kinetic energy, its momentum is left behind and drives the flow. As discussed in sections~\ref{mass} and~\ref{speed}, flow speeds in BCGs on average are similar to flow speeds in star-forming galaxies and lower than AGNs. However,  higher molecular gas masses are often found in BCGs.  Therefore, their flow momenta are higher on average compared to AGN, composite or starburst galaxies (see figure~\ref{mom-hist}). A broad range of flow momenta are found in AGNs, lying between 10$^{46}$--10$^{50}$ g cm s$^{-1}$. The flow momenta in BCGs lies between 10$^{49}$--10$^{50}$ g cm s$^{-1}$ with an average of 2.3$\times$10$^{50}$ g cm s$^{-1}$. This figure is five to ten times larger than in AGN, composite and starburst galaxies whose average flow momenta are 6.6$\times$10$^{49}$, 1$\times$10$^{49}$ and 1.7$\times$10$^{49}$ g cm s$^{-1}$, respectively. A small fraction of AGN galaxies also have a high flow momentum. 

	\section{Gas Flows and Mechanical Power}

	BCGs often host powerful radio/mechanical AGN. The mechanical power output of the AGN is estimated as the power required to inflate the buoyantly rising bubbles of relativistic plasma that are fed through jets from the vicinity of the central SMBH \citep{mcnamara07}. These bubbles are observed as cavities in X-ray images. X-ray cavities are detected in all sources in our sample. The energy required to inflate a bubble is given by $E=4PV$, where $P$ and $V$ are pressure and volume of the bubble, respectively, where the bubble is approximately in pressure equilibrium with the surrounding ICM \citep{churazov00}. The mechanical power of the AGN ($L_{\rm mech}$) can be obtained by dividing the energy by the buoyancy time of the bubble. Cavity power measurements were taken from the literature \citep{birzan12,hl15,mcdonald15,vantyghem18,calzadilla19}.

\begin{figure}
	    \centering
	    \includegraphics[width=0.45\textwidth]{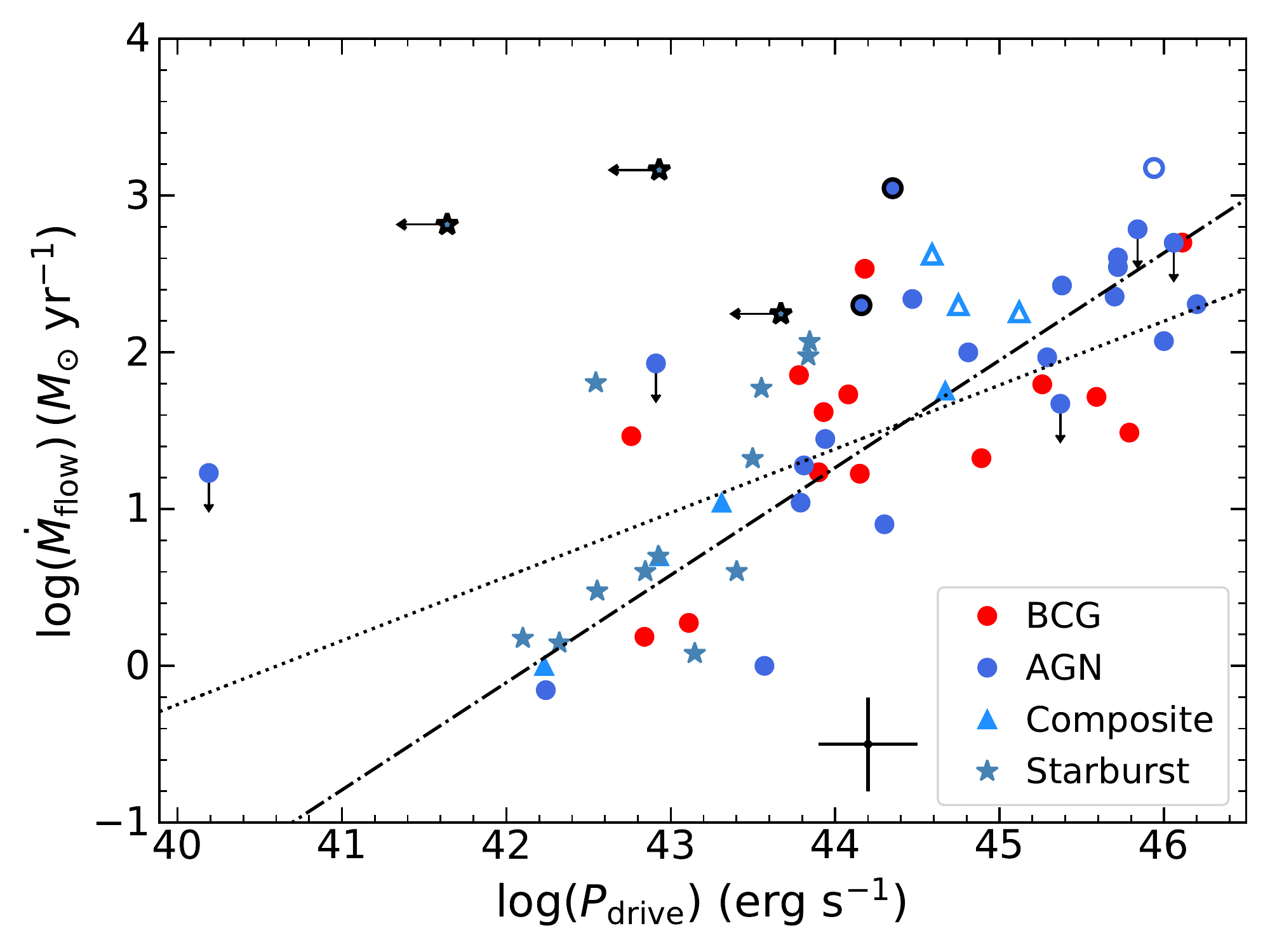}
	    \caption{The power of the driving mechanism is plotted against the molecular flow rate. The red symbols indicate BCGs. For starburst galaxies, their star formation power is used as $P_{\rm drive}$. Dark blue circles, light blue triangles and murky blue stars represent AGN, composite and starburst galaxies from \citet{fluetsch18}, respectively. The white filled symbols are flows detected in OH, and fossil flows are denoted by symbols with a black border. The dashed and dotted lines are the best fit lines for BCGs and Fluetsch galaxies, respectively.}
	    \label{M-P}
	\end{figure}

     Their mechanical powers lie between $\sim 10^{43}$--10$^{46}$ erg s$^{-1}$,
     one to two orders of magnitude larger than their nuclear radiation powers. Their mechanical powers are comparable to the radiative powers of AGN and some composite galaxies. Using the mechanical powers as $P_{\rm drive}$ for BCGs in the relationship between $P_{\rm drive}$ and $\dot{M}$ using linear regression gives a slope and intercept of 0.42$\pm$0.25 and 0.97$\pm$0.40 $M_\odot$, respectively. It is shown in Fig.~\ref{M-P} by a dotted line. The Pearson correlation coefficient 0.61 with a p-value of 0.02 indicates, again, a weak correlation. Thus, despite being comparable to the radiative powers of AGN and composite galaxies, mechanical powers are poorly correlated with the molecular gas flow rate. 
     
     This poor correlation is true for all systems including those in Fluetsch.  While trends are seen and AGN have ample power to drive the flows, the process is complex and inefficient. Perhaps most AGN power is either radiated away or it is being deposited in other forms.
     
     The ample mechanical powers and close association of molecular gas filaments with X-ray cavities in several systems suggests molecular clouds are lifted by the rising bubbles or are condensing in their updrafts. Whether this is true in all systems is not clear.  
     Evidence suggests that at least some molecular gas condenses from cooling hot gas lifted behind the X-ray cavities.  Filaments may grow due to interpenetration of hot and cold gas \citep{liu19}. Extended molecular filaments are detected towards multiple generations of cavities in the Perseus cluster \citep{salome06}.  
     However, in some instances, the mass of molecular gas is close to and may exceed the displaced atmospheric gas mass, which would be difficult to explain by uplift unless the molecular gas was created by multiple AGN cycles \citep{russell19}. The scatter in the trend between the molecular filament mass and the mechanical power indicates a complex relationship between molecular gas and the AGN. 

	\section{Lifting mechanism}
	\label{models_discus}

	\begin{figure}
		\centering
		\includegraphics[width=0.45\textwidth]{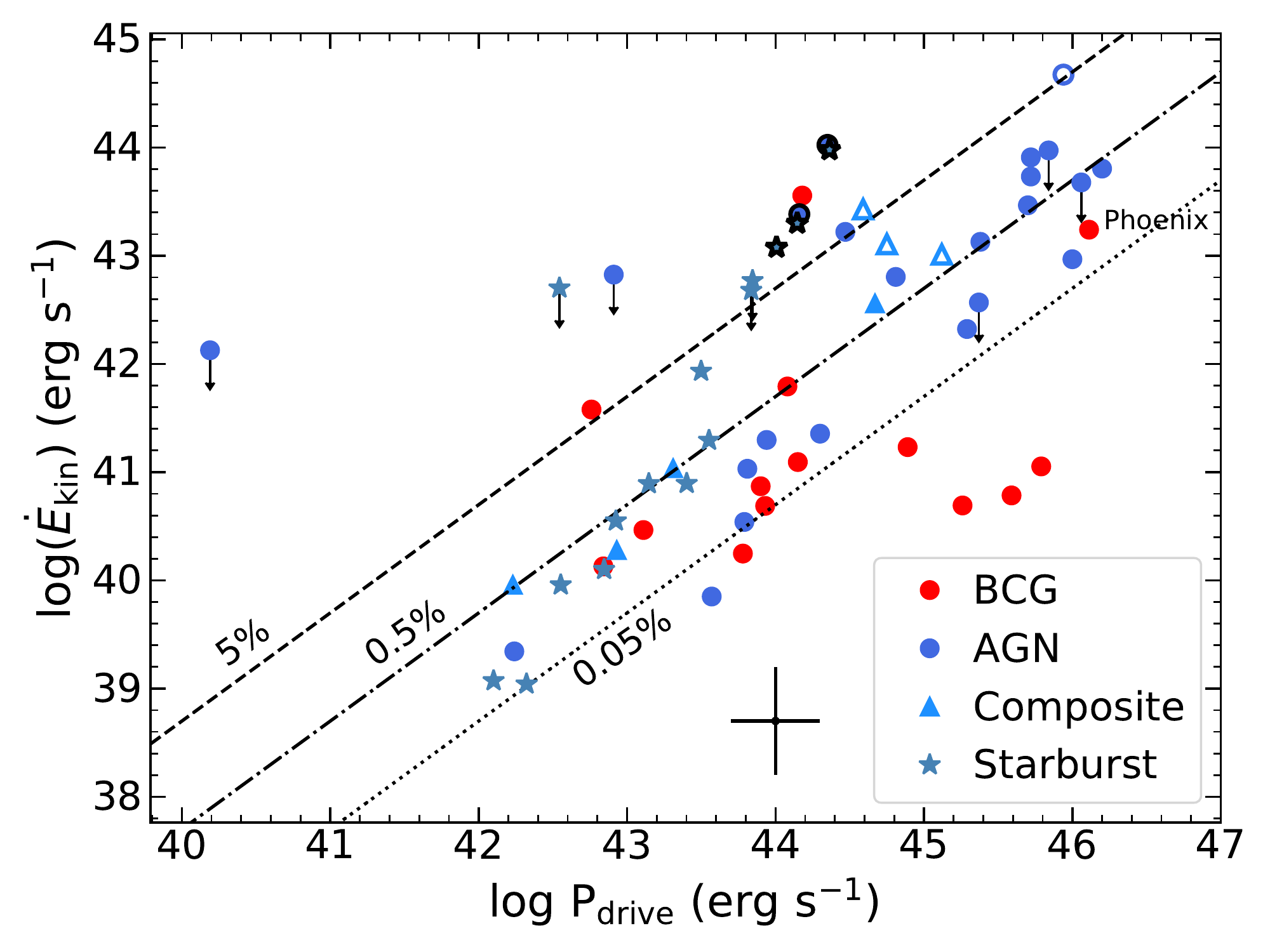}
		\caption{The figure shows the relationship between the kinetic power of flows and the mechanical powers in BCGs. For the Fluetsch galaxies, the $x$-axis represents the bolometric AGN luminosity. The black lines represent flow powers 5 percent (dashed), 0.5 percent (dashed-dotted) and 0.05 percent (dotted) that of bolometric AGN luminosity. Symbols are as in figure~\ref{M-P}.}
		\label{power}
	\end{figure}
	
	The relationship between the kinetic energy flux of the gas flow and the lifting force provides a measure of the coupling between the power source and the gas.  It is an easily-determined quantity in real and model systems and thus places interesting constraints.
	We define the molecular flow kinetic energy flux as 0.5$\dot{M} v^2$, where $\dot{M}$ is the mass flow rate and $v$ is the average speed of the flow. 
	The power for the Phoenix cluster BCG is indicated with the sum of its nuclear radiation and mechanical powers.

	Figure~\ref{power} shows that the kinetic energy fluxes of molecular gas in most systems lie near or below a few percent of the AGN or starburst power.  The radio/mechanically-driven BCGs are indistinguishable from the others in most BCGs, indicating that all mechanisms couple inefficiently to the molecular clouds and at similar levels. Some BCGs such as A1835, A2597 and PKS0745 lie much below others as a result of low flow speeds. The gas in these systems may have decoupled from the flow and slowing down. While a trend is observed in other systems, the scatter in kinetic energy flux at a given power spans 2-3 decades. Some scatter may be attributable to AGN variability.  But the radio AGN show similar scatter to nuclear AGN and starbursts.  Mechanical power is averaged over $10^7 -10^8~\rm yr$, which is comparable to the timescales for accelerating the molecular clouds.  Nuclear AGN vary on much shorter timescales \citep{schawinski15}. So the scatter probably indicates both the weak coupling and complex nature of feedback.

	The three lines in Fig~\ref{power} were chosen to reference the fractions of AGN radiation power expected in energy conserving (5\%), momentum conserving (0.5\%), and radiation pressure driven (0.05\%) flows. Most Seyfert and composite galaxies lie in the theoretically expected ranges.
	Thus, the coupling between AGN power and molecular flow kinetic energy in BCGs and other systems are broadly similar, despite very different acceleration mechanisms.

	\begin{figure}
		\centering
		\includegraphics[width=0.45\textwidth]{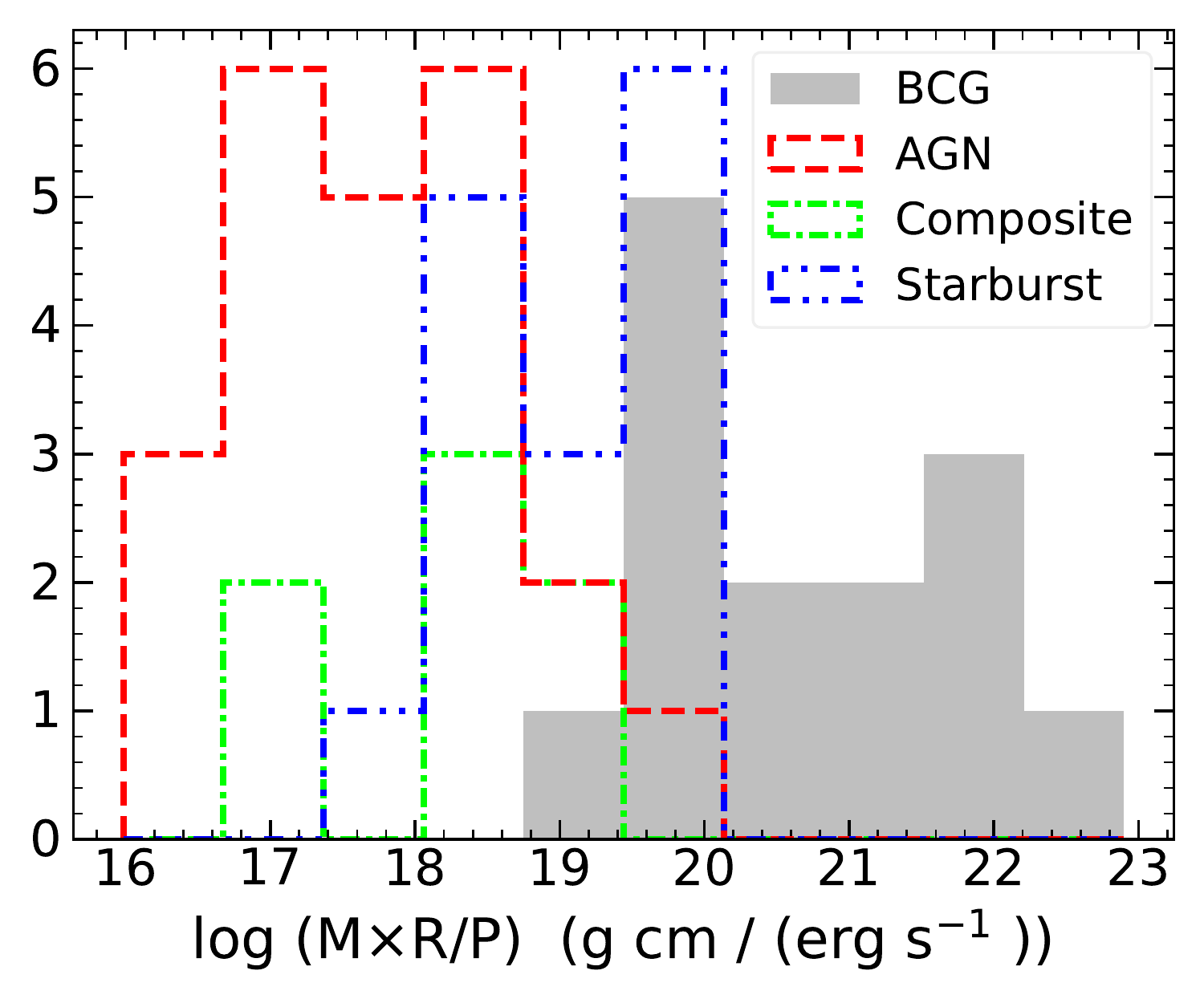}
		\caption{Histogram of flow distance times the flow mass divided by the power of the driving mechanism for BCGs, AGN, composite and starburst galaxies. This quantity is referred to as the lifting factor in the text.}
		\label{thrust}
	\end{figure}

	\subsection{Lifting Factor}
	
	Molecular gas flows in BCGs generally have higher masses and extend to larger distances on average than the systems in \citet{fluetsch18}. These properties indicate that radio AGN are generally more capable than nuclear winds, radiation, and starbursts at propelling molecular clouds out of the centers of galaxies, at least at the present epoch. 
    
    This phenomenon is explored further in Figure~\ref{thrust}, where we compare the product of the mass and flow size per unit power.  We refer to this as the lifting factor. The radio-mechanical power is adopted for BCGs. For AGN in the Fluetsch sample, we adopted the bolometric AGN luminosity. For starburst galaxies, we adopted the star formation power estimated from star formation rates using the relation $P_{\rm SFR}$ (erg s$^{-1}$) $= 2.5 \times 10^{41}$ SFR ($M_\odot$) from \citet{veilleux05}. This conversion factor is consistent with the factor measured independently in the central galaxy Abell 1835 \citep{mcnamara06} which is included in our sample.  This single example offers reassurance that the conversion relation of \citet{veilleux05} applies to BCGs.
    
    Figure~\ref{thrust} shows BCGs have higher lifting factors on average than other outflow systems,  indicating that radio-mechanical feedback is more effective per unit power at lifting large masses to greater distances than nuclear AGN and starbursts. The lifting factor for starbursts (H2 galaxies) is considerably larger than for AGNs.  The starbursts overlap the lifting factors of the weakest BCG flows, but BCGs outperform all by more than an order of magnitude using this figure of merit.
    
    Figure~\ref{thrust} can be interpreted in the context of Figure~\ref{power}. In Figure~\ref{power} we show that the molecular kinetic energy fluxes against driving power are of similar magnitude between the BCGs and galaxies in the Fluetsch sample.  At first blush, Figure~\ref{power} and Figure~\ref{thrust} appear to be inconsistent with each other.  This apparent inconsistency is attributable to the large differences in timescale and size of the flows. The AGN in Fluetsch are driving smaller masses of molecular gas over shorter distances but at higher speeds.  The much larger masses, distances and timescales in BCGs compensate such that the kinetic energy fluxes per unit power are similar. However, the total mass displaced over time is much larger in BCGs for a given mean power. 
   
    The large range of lifting factors may be due to variations in flow characteristics including the volume affected by radio, nuclear, and star-formation activity and the ability of the working surfaces to couple to the ambient gas. This coupling is always weak but varies greatly (Fig.~\ref{power}).  
    
    Radiation pressure wind energy is released on smaller scales close to the AGN, where radiation intensity is strong, and the particle number density is high.
    At larger distances the medium becomes tenuous, and the radiation intensity drops rapidly. The outflow then transitions into a momentum conserving flow and rapidly slows down \citep{veilleux20}. Therefore, radiation driven flows are less efficient at lifting a large amount of mass to vast distances. 
    
    The large lifting factor of radio-mechanical feedback in BCGs is attributable to the large volume of impact and the relatively long timescales radio bubbles are able to lift gas in the surrounding interstellar medium and atmosphere.   Radio bubbles couple to the gas in the inner kpc \citep{mukherjee18,guo18} lifting the low entropy gas in their wakes to high altitudes through drift and entrainment \citep{pope10}. This is observed in real systems as high metallicity atmospheric gas located in the wakes of rising X-ray bubbles extending in some instances to altitudes of tens of kpc \citep{kirkpatrick15}. 
    
    Radio bubbles encompass a large range of sizes and volumes, with diameters of a few kpc to over 200 kpc.  A typical bubble appearing as an X-ray cavity is elliptical in projection with an average semi-major and semi-minor axis of 11.4 kpc and 7.7 kpc, respectively \citep{rafferty06,calzadilla19,mcdonald15,hl15,vantyghem17}. Assuming ellipsoidal 3D shape, the average volume of a bubble is 10$^{68}$ cm$^3$.  In contrast, the average volume of the AGN wind-driven outflows is $\sim 10^{65}$ cm$^3$, assuming cylindrical geometry with a radius and height of $\sim$1 kpc, respectively. Starbursts typically occur within the central 1 kpc. The energy injection region is typically up to $\sim$2 kpc in extreme situations, after which the flow starts slowing down \citep{schneider20}.  Thus, the effected volume is approximately 16$^{65}$--10$^{66}$ cm$^{-3}$. The volume would be smaller considering a more realistic bi-conical geometry.
    Therefore the working volume of radio-mechanical feedback is much larger and affects a much larger mass for a given ambient gas density.

    Another key factor is the long lifting timescales of radio/X-ray bubbles. Their typical observed ages lie between 10--20 million years \citep{birzan04}.  But many survive for $ \sim $ 100 million years or longer \citep{brienza21}.  While the jet launching phase typically lasts for $\sim$10 million years, comparable to a typical quasar lifetime \citep{martini03,bird08,schawinski15}, radio bubbles propelled outward by buoyancy continue to draw in and lift gas in their updrafts.  Therefore, a single radio event can continue to drive gas outward long after a quasar of similar power shuts down. Once it shuts down the gas slows down in the galaxy's gravitational potential and by ram pressure and drag forces. 
    
    We have considered here only the displaced molecular gas in BCGs. The effects shown in Figure~\ref{thrust} are likely larger. Rising radio bubbles also lift and displace hot gas from the central atmospheres. $Chandra$ has observed metal enriched gas preferentially along the radio jet axes \citep{kirkpatrick11,simionescu08,gitti11}. Similar features are reproduced in hydrodynamic simulations indicating 10$^9$--10$^{10}$ M$_\odot$ of metal rich gas lifted from central regions of galaxy clusters \citep{duan18,qiu20,li15}. 
    
    The atmospheric mass displaced by cavities is comparable to or larger than the molecular gas masses \citep{russell19}. \citet{fluetsch18} found the molecular masses in their sample exceeded the atomic and neutral phases by roughly 40 times. Therefore, in order to compare our systems with Fluetsch, we have not included the hot gas masses.  But were we to do so the flow rates and masses would increase by a factor of two or more.

\begin{figure}
		\centering
		\includegraphics[width=0.45\textwidth]{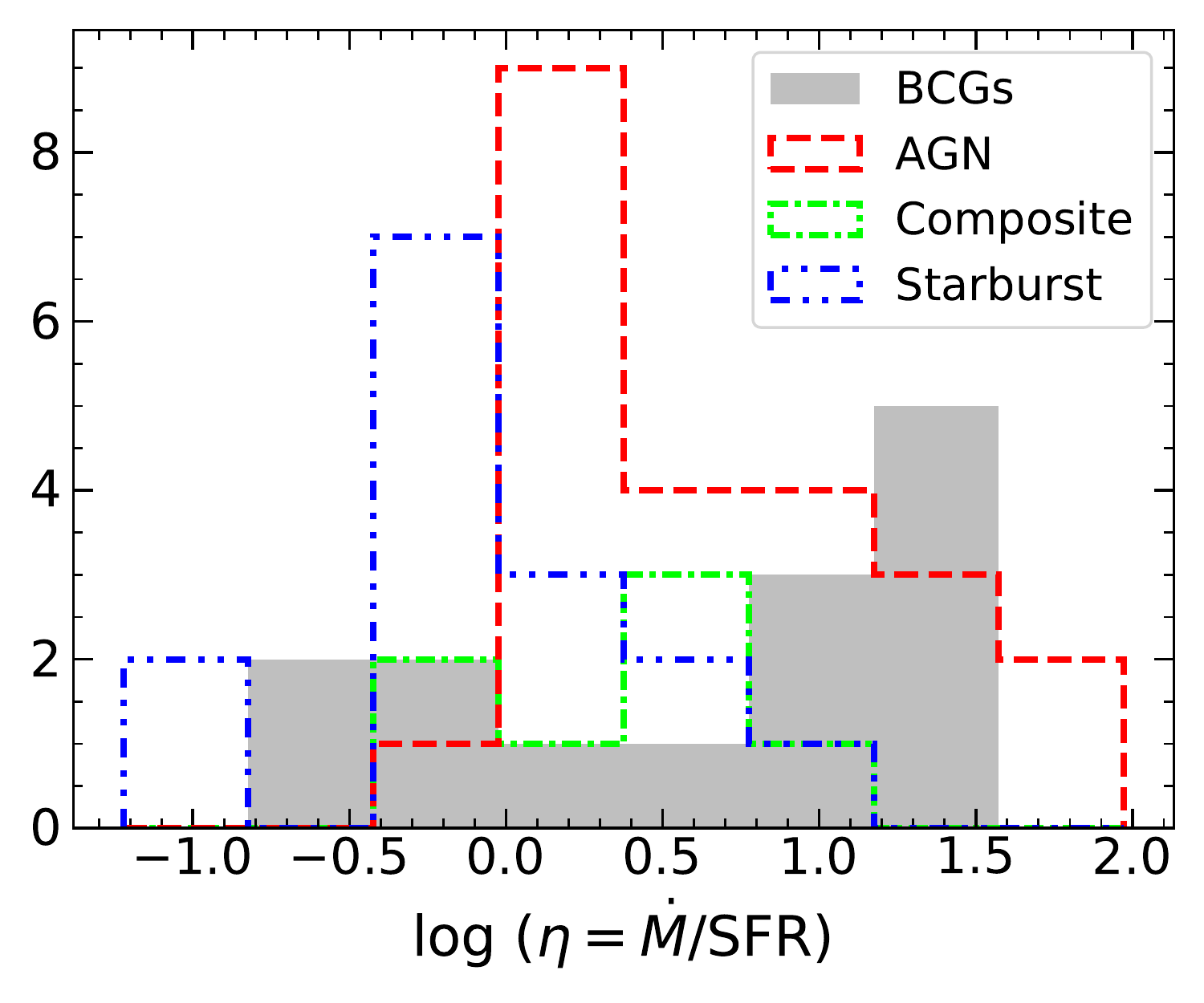}
		\caption{Histogram of log of mass loading factor in BCGs, AGNs, composite and starburst galaxies.}
		\label{loadingfact}
\end{figure}
		
\section{Star Formation}

Here we explore relationships between feedback, outflows, and star formation.
We adopt BCG star formation rates (SFR) from \cite{mcdonald18} who used several methods to estimate the SFRs.  For Pheonix, they decomposed the spectrum to separate the AGN component from stellar radiation. For the remaining objects, they adopted the logarithmic mean of multiple SFRs for each system found from the literature. The average logarithmic scatter in measurements of SFRs for BCGs is determined to be 0.28 dex. The SFR measurement for RXCJ0821 is taken from \cite{odea08}.

	\subsection{Mass loading factor}
	\label{masslf}
	The mass loading factor $\eta$ is defined as the ratio of the molecular gas flow rate to the star formation rate ($\eta = \dot{M}/SFR$). A high value of $\eta$ indicates that the AGN can sweep the gas from the galaxy as quickly as star formation consumes it. Low $\eta$ indicates that star formation will consume a significant amount of gas before it can be removed from the inner regions of the galaxy. The histogram in figure~\ref{loadingfact} shows the distribution of $\eta$ for different galaxy types, including the fossil galaxies in the Fluetsch sample. The majority of the star-forming galaxies have mass loading factors  $\leq$1. This is expected from feedback models for star-forming galaxies in which supernova explosions are the primary outflow driving mechanism. The galaxies hosting AGN from the Fluetsch sample and BCGs have a broad range of $\eta$ from 0.1 to $\sim$100. However, the majority of those galaxies have $\eta >1$.
	
    High values of $\eta$ found preferentially in BCGs and AGN galaxies indicate that much gas is displaced from these systems before it can form stars. Therefore, star formation may be suppressed or quenched assuming the gas leaves the galaxy and does not return. In the next section we show that very little molecular gas leaves these systems.

	\subsection{Does Radio-mode Feedback Suppress Star Formation?}
	
	\begin{figure}
		\centering
		\includegraphics[width=0.5\textwidth]{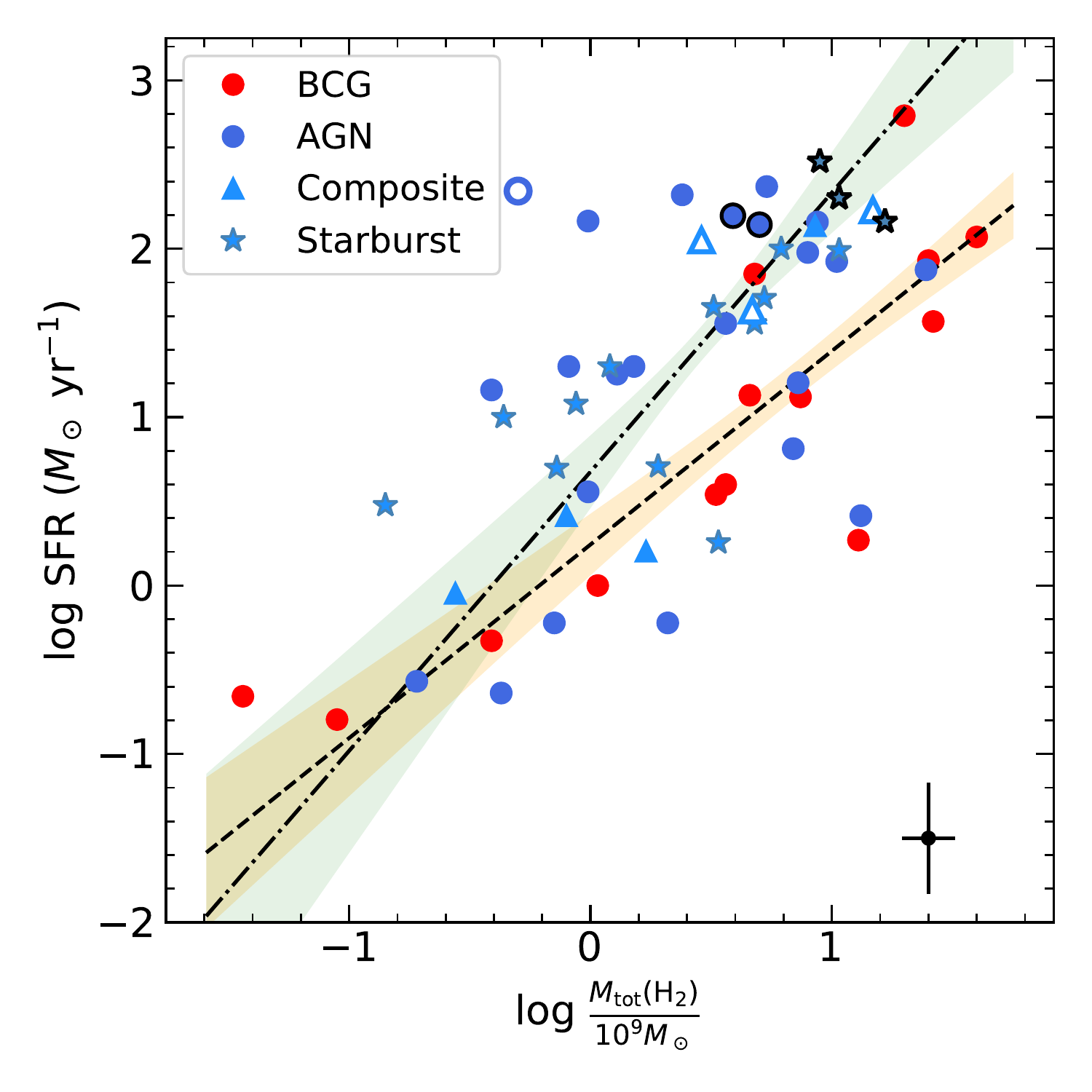}
		\caption{The figure shows the total molecular gas mass plotted against the SFR of host galaxies. The dashed line and orange region show the best-fit line and 1$\sigma$ confidence interval for BCGs, respectively. The dashed-dotted line and the green shaded region show the same for all Fluetsch galaxies. All symbols are as in figure~\ref{M-P}.}
		\label{sfr-m}
	\end{figure}
	
	Figure~\ref{sfr-m} shows the relationship between the total molecular gas mass and the star formation rate for BCGs and the Fluetsch galaxies.  Two clear trends are seen in Figure~\ref{sfr-m}.  Linear regressions of the form log SFR ($M_\odot$ yr$^{-1}$) = $a$ log $\frac{M_{\rm tot}}{10^9 M_\odot}$ $+$ $b$ were fitted separately to the data for the Fluetsch galaxies and BCGs.  The values of parameters ($a$, $b$, $\Delta a$, $\Delta b$) for BCGs and Fluetsch galaxies are (1.13, 0.23, 0.26, 0.24) and (1.66, 0.67, 0.43, 0.23), respectively.  $\Delta a$ and $\Delta b$ are the 1$\sigma$ errors. The best fit-lines for BCGs and Fluetsch galaxies are shown by the dashed and dashed-dotted lines, respectively. 
	
	The values of the parameter $a$ indicate the two populations have different slopes. BCGs have a lower total star formation rate per unit molecular gas mass compared to galaxies from the Fluetsch sample.  The segregation becomes strikingly apparent for total molecular gas masses greater than $\sim$10$^9$ $M_{\odot}$. To test the statistical significance of the difference observed between the two best-fit lines, we conducted a two-sample t-test with a null hypothesis that the coefficients of the two linear regressions are equal. The t-statistic for the slope is 5.6 with a p-value smaller than 0.005. Similarly, the t-statistic for the intercept is 5.5 with a p-value smaller than 0.005. Therefore, we can confidently reject the null hypothesis. The observed difference between BCGs and Fluetsch galaxies in the relationship between their SFR and $M_{\rm tot}$ in log space is statistically significant.
	Given the small sample size, we cannot determine whether the separate trends are due to selection bias.
	
	
	
	To further investigate these potentially interesting trends, we compared the star formation in BCGs and galaxies from Fluetsch with the Kennicutt-Schmidt (KS) relation. KS relates the star formation rate surface density ($\Sigma_{\rm SFR}$) to the total (H I+H$_2$) cold gas surface density \citep{kennicutt98}.  The KS relation is well-characterized, fits a broad range of galaxy classes, and scales non-linearly as $\Sigma_{\rm SFR}$ $\propto$ $\Sigma_{\rm H I+H_2}^{1.4}$. 
	
	We compared the SF law in BCGs and Fluetsch galaxies with the KS relation. We adopted star formation rates in BCGs listed in Table~\ref{tab:bcg_properties}, and the remaining star formation rates were taken from \citet{fluetsch18}. In most Fluetsch galaxies, star formation occurs in the circumnuclear region within $\sim$1 kpc. Star formation rates in galactic disks typically lie between 1--20 $M_\odot$ yr$^{-1}$ \citep{kennicutt98}. Therefore, we have adopted the area of a 1 kpc radius region around the nucleus to estimate gas and star formation surface densities for Fluetsch galaxies with star formation rates greater than 20 $M_\odot$ yr$^{-1}$. For the remaining galaxies from the Fluetsch sample, we obtained their angular diameters from NED\footnote{\href{https://ned.ipac.caltech.edu/}{https://ned.ipac.caltech.edu/}}. We used blue band diameters from the RC3 - Third Reference Catalog of Bright Galaxies \citep{corwin94} and estimated areas assuming elliptical shape. In eight BCGs, the areas of the star formation regions were estimated using UV-band images from the Hubble Telescope. The star formation regions in these BCGs are confined within roughly 3 kpc to 10 kpc of the center. Star formation in Phoenix and Perseus extend up to $\sim$25 kpc in filamentary structures. For the remaining BCGs, we adopted a conservative area of the inner 3 kpc region centered on the nucleus. In some BCGs (e.g., Abell S1101, RXCJ1539.5, PKS0745), molecular gas is more extended than the star forming region. Therefore, the molecular gas surface densities may be overestimated by a factor of $\sim$2 in these systems.
	
\begin{figure}
	    \centering
	    \includegraphics[width=0.5\textwidth]{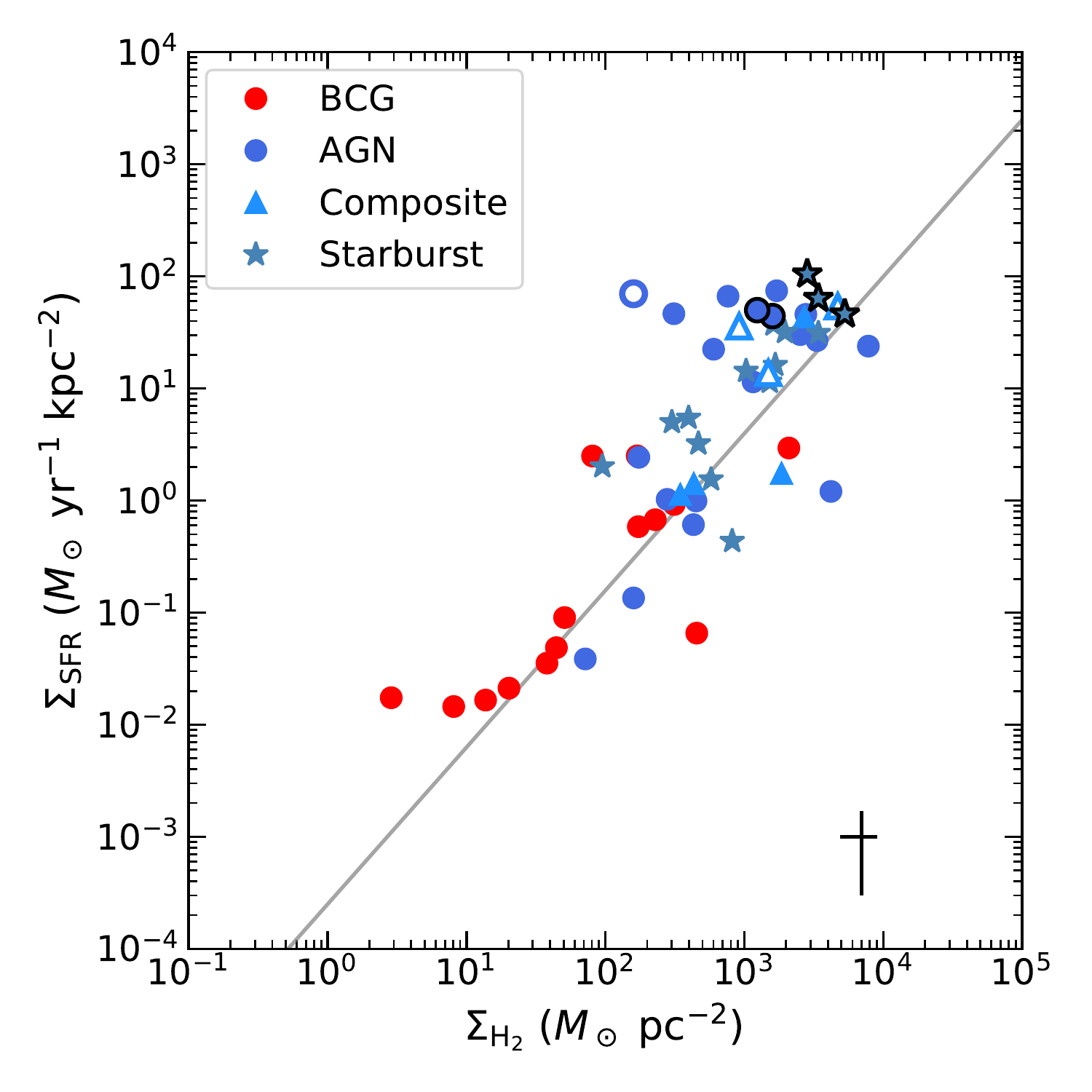}
	    \caption{The figure expresses the relationship between the molecular gas surface density and the SFR surface density for BCGs and FL galaxies. The grey line shows the global Kennicutt-Schmidt relation \citep{kennicutt98}. Symbols are as in figure~\ref{M-P}.}
	    \label{fig:ks}
	\end{figure}

	Figure~\ref{fig:ks} shows the star formation law in BCGs and Fluetsch galaxies when only molecular gas surface density is used for all galaxies. The best-fit relation for all galaxies is given by:
	\begin{equation}
	\Sigma_{\rm SFR} = 10^{-3.09\pm0.49} \Sigma_{\rm H_2}^{1.34\pm0.17},
	\end{equation}
	where, $\Sigma_{\rm SFR}$ is in $M_\odot$ yr$^{-1}$ kpc$^{-2}$ and $\Sigma_{\rm H_2}$ is molecular gas surface density in $M_\odot$ pc$^{-2}$. The overall distribution appears consistent with the KS-relationship. The dispersion around the KS relation is larger than the observational uncertainties alone, implying real variance from galaxy to galaxy. Some of this variance may be attributed to variations in the $X_{\rm CO}$ factor used to convert CO surface brightness to molecular gas mass \citep[see][for review, and references therein]{kennicutt12}. Most BCGs fall into the intermediate density star formation regime occupied by normal spiral disks galaxies and the inner regions of the Milky Way. The star formation surface densities of most Fluetsch galaxies are higher than the BCGs and are typical of those observed in circumnuclear starbursts and ULIRGs. This is to be expected as the Fluetsch sample contains several ULIRG and starburst systems. 
	
	The Fluetsch galaxies lie systematically above the KS relation, such that they have higher than normal star-formation-rate surface densities compared to their molecular gas mass surface densities. Whether this departure is real or due to measurement bias is unclear.  The KS relation is properly characterized as the sum of the atomic and molecular gas surface densities. We consider here only the molecular gas mass because H I emission is rarely observed in BCGs, whose cold gas reservoirs are likely dominated by molecular gas with H$_2$/H I ratio unity or above \citep{babyk21}.
		
	
	Why BCGs have lower star formation rates per molecular gas mass compared to the Fluetsch galaxies is unclear.  The location of the Fluetsch galaxies above the KS law while the BCGs follow the law suggests the Fluetsch galaxies have higher than normal star formation rate densities.  The BCGs are outwardly normal.  This difference may be related to the fact that half of the total molecular gas in BCGs is found in off-nuclear filaments and is rarely seen in circumnuclear disks \citep{russell19,olivares19}. The filaments, being dynamically disordered and perhaps younger than disk gas, may be less prone to gravitational instabilities leading to star formation.  The magnetic field may further reduce fragmentation.  Mixing with hot surrounding gas and magnetic reconnection may reheat the cold filaments before star formation can ensue \citep{fabian11, churazov13}. The gas in the nuclear regions is also dynamically disturbed in many BCGs. The lack of ordered structure in the nucleus suggests that the gas is dynamically young or is continually destroyed and reformed within the central galaxy, preventing it from being consumed by star formation on short timescales. Perhaps one more of these factors reduces the efficiency of star formation per unit local free-fall time in BCGs compared to AGN, composite and starburst galaxies. These suggestions are speculative and merit rigorous treatment in the future.

	\section{Fate of the Outflowing Gas}
	\label{escape}
   
   The final disposition of the outflowing gas depends on many factors including, the loading factor, the speed of the flow relative to the escape speed, and the rates of formation and destruction of molecular clouds.  
   Figure~\ref{loadingfact} indicates broad similarity across all feedback modes in nearby AGN, albeit with a large scatter. Loading factors near and above unity indicate that winds are as effective or more effective than star formation at depleting the gas supply.  
   
   The degree to which star formation is delayed or suppressed depends largely on whether the molecular gas escapes the galaxy.  Fluetsch found that in systems with the highest outflow speeds, only a few percent of the molecular gas escapes entirely.  In most instances escape is negligible. If the currently outflowing gas returns to the galaxy in molecular form, it would be available for future star formation.  This scenario is almost certainly true in BCGs, where molecular gas velocities are low compared to both the free-fall speeds and escape speeds \citep{russell17a,russell17b,russell19}. This is compounded by ram pressure forces on the gas by the surrounding hot atmosphere.  The emerging picture is a fountain where hot and possibly cold molecular gas is lifted behind the bubbles.  The gas eventually cools.  Most will likely return to the galaxy, but some molecular gas may be destroyed.  The details are unclear.

	To compare the molecular gas speeds to the escape speeds, we adopt the Hernquist profile to model a galaxy's gravitational potential.  The estimated escape velocities are found using the following formula:
	\begin{equation}
	v_{\rm esc} = \sqrt{\frac{2GM}{r+a}}
	\end{equation}
	where $a$ is the scale radius, which is related to the effective radius as $R_e \sim 1.8153a$, and $M$ is the stellar mass of the galaxy taken from table~\ref{tab:bcg_properties}. To obtain an approximation to the escape velocity for the  Phoenix cluster BCG, we adopt the value 17 kpc from \citet{mcdonald12a}.  
	For the remaining galaxies, we use an empirical relation between effective radius and stellar mass for local elliptical galaxies from \citet{mcintosh05} to determine $R_e$. This method overestimates the effective radius for Phoenix by $\sim$65 percent. Therefore, we adjusted effective radii to take this factor into account. The escaped gas was defined as the molecular gas with a velocity greater than the escape velocity of the galaxy. To estimate the escaped gas fraction, we integrated the spectrum above the positive and below negative escape velocity and compared it to the total integrated intensity of the flow. 
	
	We found that for all BCGs the escaped gas fraction is negligible. The high-velocity gas in Abell 1664 has sufficient velocity to escape the central region of the BCG. But it lies well below the escape velocity of the central cluster halo and will likely return.
	
	As the gas cannot leave the galaxy, it will likely return. Thus some of the molecular gas in BCGs is almost certainly flowing inward. Discussed in section~\ref{size}, it is extremely difficult to distinguish an outflow from an inflow with gas seen in emission.  However, molecular gas is seen in absorption against the nuclear continuum of some systems has revealed both in- and out-bound molecular gas with velocities between $-45~\rm km~s^{-1}$ and $283~\rm km~s^{-1}$ \citep[][]{david14,tremblay16,rose19}. The absorption velocities are broadly consistent with the molecular fountain model.

	\section{Concluding Remarks}
    
    We have shown that mechanical energy released by radio jets can have a far more dramatic impact on galaxies compared to nuclear radiation (QSOs) and winds in contemporary active galaxies. This is also true in ancient massive galaxies lying at high redshift.  The relative impact of radio-mechanical feedback, characterized here as the lifting factor, is qualitatively similar to the situation in high redshift radio galaxies experiencing both quasar and radio-mechanical feedback.  In a study of 24 radio galaxies lying at $z\sim 2$, \citet{nesvadba17} examined the energy and momentum imparted on the surrounding ionized gas by star formation, radio jets, and nuclear AGN.  They found that while starburst winds play a minor role, radio jets are generally more effective than quasars at powering gas motions. 
    
    Likewise, \citet{kellerman16} found that 20\% of contemporary quasars 
    lying between $z = 0.2-0.3$ are radio-loud, double-lobed systems, akin to those at higher redshift.  Furthermore, \citet{jarvis19} noted that compact radio jets with relatively modest radio luminosities hosted by some quasars lying within the \citet{kellerman16} redshift range are interacting with and driving outflows of nebular gas.  Assuming radio jets are operational during 20\% of the life of a quasar, as the \citet{kellerman16} results may imply, their large lifting factors indicate that radio-mechanical feedback would have a significant effect on the evolution of their host galaxies.
    
    
    A difference between the Nesvadba study and ours apart from epoch is the ability to examine the three forms of feedback occurring simultaneously in the same galaxies. While this approach is rarely possible in the contemporary Universe, some central cluster galaxies are simultaneously experiencing powerful nuclear AGN, starbursts, and radio-mechanical feedback.  For example,  IRAS 09104, is a radio galaxy with large X-ray cavities surrounding a quasar host lying at $z=0.44$.  Its quasar power exceeds its radio-mechanical cavity power by more than an order of magnitude. Yet its cavities are driving $\sim 4.5 \times 10^{10}~\rm M_\odot$ of molecular gas out of the central galaxy with little help from the quasar \citep[see][]{osullivan21}.  The same is true for the iconic Phoenix cluster central galaxy \citep[see][]{russell17a} where radio-mechanical feedback is the primary source of energy and momentum input to its $\sim 10^{10} ~\rm M_\odot$ well of molecular gas. We now understand that this feedback mechanism has been operating effectively 
    through much of the history of central cluster galaxies \citep{ma13, hl15} and a broader spectrum of massive galaxies \citep{nesvadba17}.

 Powerful radio AGN are found commonly in massive giant ellipticals \citep{best03,best05,heckman14} often hosting dense, high pressure atmospheres.  Radio AGN are also found in giant ellipticals in less dense atmospheres capable of driving both HI and molecular flows \citep{morganti05}. In these systems the radio energy may be a significant fraction of the thermal energy in their atmospheres, indicating that the radio jets are heating the atmospheres and driving away hot gas from their elliptical galaxy hosts \citep{webster21b, morganti21}.

How far down the radio luminosity function radio-mode feedback is affecting galaxy evolution is unknown. Radio surveys such as NVSS and First detect primarily low radio luminosity galaxies \citep{tadhunter16}. But even low power radio galaxies trace much higher mechanical powers \citep{birzan08,croston18} and presumably higher lifting factors. Recent observations suggest that the mechanical feedback could be significant in 10\% of Seyfert galaxies \citep[for example,][]{webster21a} whose hosts are spiral rather than elliptical galaxies. Even dwarf galaxies experience AGN feedback \citep{mk19}, some driven by mechanically-powerful radio jets \citep{mezcua19,davis22}. 
 
The upshot here is that when active, radio jets  can strongly influence the evolution of galaxies from their nascency to mature contemporary galaxies.  
The systems studied here do not represent the general population of galaxies.  Instead, they are snapshots of an active period through which most massive galaxies transit.
It would be premature to draw broad conclusions about the influence of radio-mode feedback on the general population of galaxies and their nuclear black holes.
This must await further investigations of large samples of galaxies observed across the electromagnetic spectrum from radio to X-rays.

	\section{Summary}
	Molecular gas properties are examined in 14 active galaxies (BCGs) centered in clusters with cooling atmospheres. The molecular gas properties and power output from AGN and star formation were compared to 45 local active galaxies compiled by \citet{fluetsch18}.  Our results are summarized as follows:
	
	\bigskip
	\noindent
	$\bullet$ BCGs centered in cooling atmospheres contain $\sim 10^8$ $M_\odot$ to upwards of 10$^{10}$ $M_\odot$ of molecular gas.  Thirty to seventy percent of the gas lies outside of the nucleus in extended, filamentary structures that appear to be moving relative to the BCG. Gravitationally stable structures, such as large-scale disks, are rare \citep{olivares19,russell19}.  In contrast, only a few to ten percent of the total molecular gas mass in contemporary AGN and starburst galaxies examined by \citet{fluetsch18} is flowing inward or outward. The remainder presumably lies in disks or other stable structures.
	
	\bigskip
	\noindent
	$\bullet$ Molecular clouds surround or lie beneath X-ray cavities inflated by radio jets in many systems. The clear association of molecular clouds with X-ray cavities in systems such as the Perseus cluster (NGC 1275), the Phoenix cluster, Abell 1835 and others indicate that molecular clouds are being lifted outward, clearing the nucleus of gas. Molecular clouds may also be condensing from atmospheric gas lifted in the updrafts of the rising radio bubbles (cavities).  The molecular mass lifted by radio bubbles can
	exceed the masses of gas flows studied by \citet{fluetsch18} by factors of 10--100. 
   
    \bigskip 
   \noindent 
    $\bullet$   
     Radio bubbles are able to lift molecular material to altitudes of $\sim 10$ kpc and beyond, with flow sizes on average $\sim$10 times larger compared to active galaxies compiled in \citet{fluetsch18}.  Cluster outflows tend toward lower velocities than those in the comparison sample. However, their momentum fluxes are, on average, an order of magnitude larger.  
    
    \bigskip 
   \noindent 
    $\bullet$  
     We introduce the lifting factor, a parameter that is the product of the mass and size of the molecular flow divided by the driving power. The driving power is characterized by the measured AGN power or the starburst power. This parameter indicates that radio-mode feedback is vastly more capable of driving large gas masses to higher altitudes per unit driving power than other active systems. This result is consistent with similar measurements for powerful radio galaxies at redshifts beyond 2 (Section 11).

     \bigskip
     \noindent
     $\bullet$ Loading factors vary broadly in BCGs from $\sim 0.1-100$.  This range is similar to the active galaxies compiled by \citet{fluetsch18}. Only a few percent at most of the AGN mechanical power in BCGs is transmitted to the molecular gas.  This fraction is also similar to the systems in \citet{fluetsch18}. 
    
    \bigskip
    \noindent
     $\bullet$  The star formation rate per gram of molecular gas in BCGs is five to ten times lower than the most massive systems with AGNs and starbursts studied by \citet{fluetsch18}. As molecular gas is closely associated with star formation, this deficit indicates, tentatively, that star formation in BCGs is suppressed compared to other systems.
     
   \bigskip
   \noindent
    $\bullet$
     Molecular cloud velocities in BCGs lie well below escape speeds. Similar to the systems in \citet{fluetsch18}, little or no molecular gas is able to escape the galaxy. Radio-mechanical feedback is likely driving a fountain of hot and cold gas that eventually returns to the central galaxy.  This process may delay or suppresses star formation relative to other active galaxies.

  \bigskip
    Radio-mechanical feedback (radio-mode) is more complex than is commonly assumed. Radio sources not only heat their surrounding atmospheres but also drive molecular clouds out of their host galaxies.  In some instances, radio-mode feedback may promote the condensation of molecular clouds from their hot atmospheres that would sustain long-term feedback needed to prevent the outsized growth of galaxies. Radio lobes are more capable of driving large masses of molecular gas to higher altitudes than contemporary starbursts and quasars.  These effects may be most prominent in dense cluster atmospheres where the cooling gas supply is plentiful and buoyancy forces are large.  Radio/X-ray bubbles encompass large volumes of gas that vastly exceed other AGN, where buoyancy effectively drives gas outward long after the AGN has ceased to power the radio lobes.  
    
    Understanding when molecular clouds are inbound or outbound is fraught with uncertainty. That most molecular gas lies off the nucleus shows much of it is outbound. But because their speeds lie well below the escape speed, some of the gas must be inbound.  Our results will thus be biased to some degree. But quantities such as the lifted mass will be affected by no more than a factor of two and will not qualitatively affect our conclusions.  Lacking a complete sample, we are subject to selection bias, which likely will affect properties of lower molecular gas mass systems that are undersampled.  Selection biases will be addressed as sample sizes increase and unbiased samples become available. 

\section*{Acknowledgements}

BRM acknowledges support from the Natural Sciences and Engineering Council of Canada. ACE acknowledges support from STFC grant ST/P00541/1. HRR acknowledges support from an STFC Ernest Rutherford Fellowship and an Anne McLaren Fellowship. We thank the referee for their comments which significantly improved the manuscript. This paper makes use of the following ALMA data: ADS/JAO.ALMA\#2012.1.00837.S, ADS/JAO.ALMA\#2011.0.00374.S, ADS/JAO.ALMA\#2015.1.00623.S, ADS/JAO.ALMA\#2012.1.00988.S, ADS/JAO.ALMA\#2012.1.00837.S, ADS/JAO.ALMA\#2013.1.01302.S, ADS/JAO.ALMA\#2016.1.01269.S, ADS/JAO.ALMA\#2011.0.00735.S, ADS/JAO.ALMA\#2015.1.01198.S. The National Radio Astronomy Observatory is a facility of the National Science Foundation operated under cooperative agreement by Associated Universities, Inc. The scientific results reported in this article are based in part on data obtained from the Chandra Data Archive.

This research made use of {\sc python} \citep{rossum09}, {\sc Astropy} \citep{astropy13,astropy18}, {\sc matplotlib} \citep{hunter07}, {\sc numpy} \citep{walt11,harris20}, and {\sc scipy} \citep{scipy20}. We thank their developers for maintaining them and making them freely available.

\section*{Data Availability}

The ALMA data used in this article is publicly available at \href{
https://almascience.nrao.edu/aq/}{https://almascience.nrao.edu/aq/}.



\bibliographystyle{mnras}
\bibliography{ref} 




\appendix

\section{Description of targets}
\label{appA}
	
	The molecular gas distribution in BCGs is often clumpy and asymmetric. Filaments exist in most systems with narrow velocity widths and smooth gradients. Filaments are often found behind rising X-ray cavities or encasing the cavities. The fraction of molecular gas in filaments ranges from greater than 70 per cent to greater than 30 per cent. Most systems exhibit multiple velocity components in the molecular gas spectra. The spatial extent of the filaments determines the size of the flow. In the following subsections, we briefly describe molecular gas morphology and kinematics in each target in our sample.
	
	\begin{figure*}
		\centering
		\includegraphics[height=0.9\textheight]{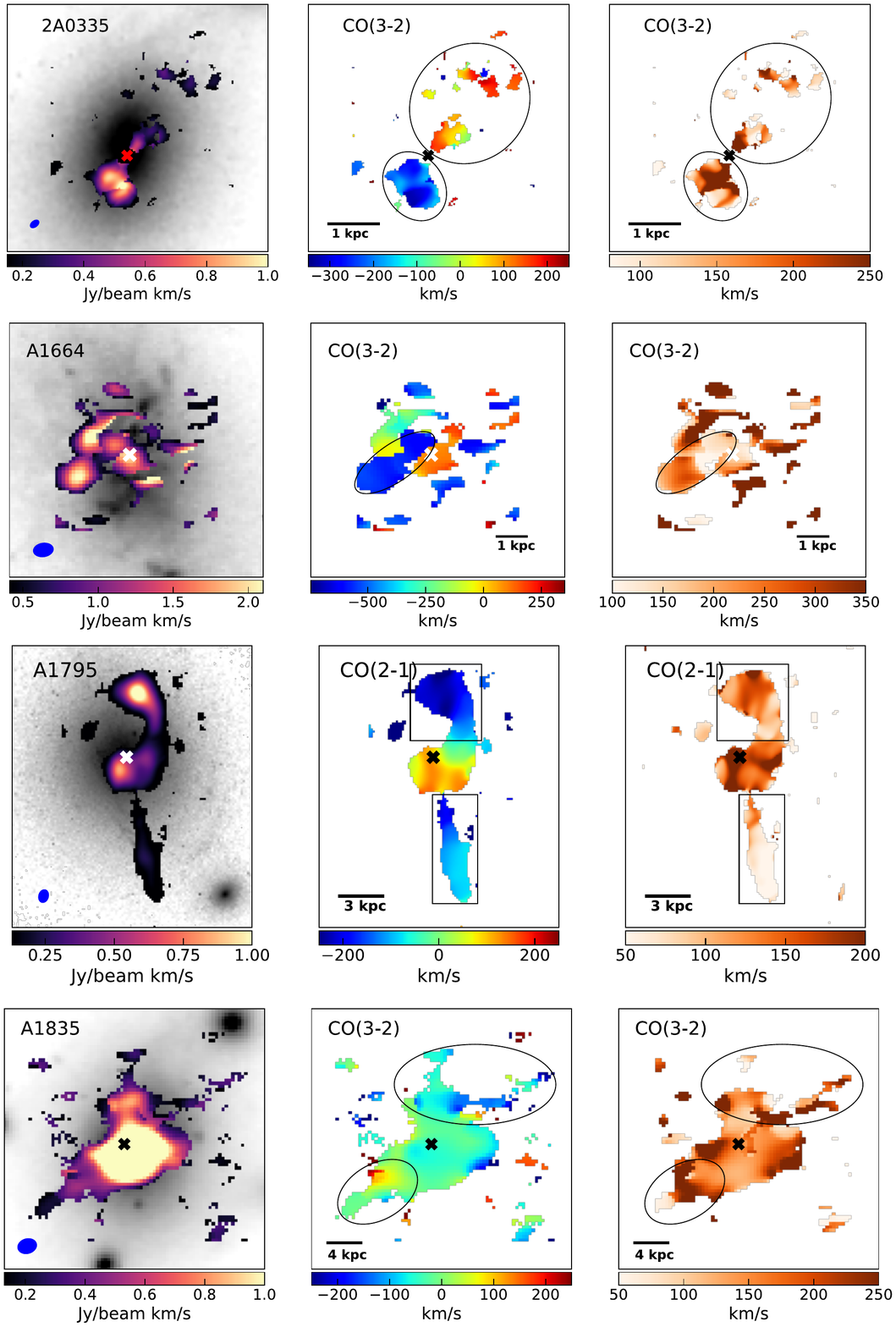}
		\caption{ALMA CO integrated intensity map overlaid on HST images (left column), velocity centroid maps (middle column), and FWHM maps (right column) for all BCGs in our sample are shown. For RXCJ1539,  DSS red band image is shown. The beam size is indicated as a grey ellipse. The crosses indicate the location of the AGN in each BCG. The emission region enclosed by black ellipses or polygons in the right panels of each row indicates the region we consider as a flow in each BCG.}
		\label{maps}
	\end{figure*}
	\renewcommand{\thefigure}{A\arabic{figure} (Continued)}
    \addtocounter{figure}{-1}
	\begin{figure*}
	    \centering
		\includegraphics[height=0.9\textheight]{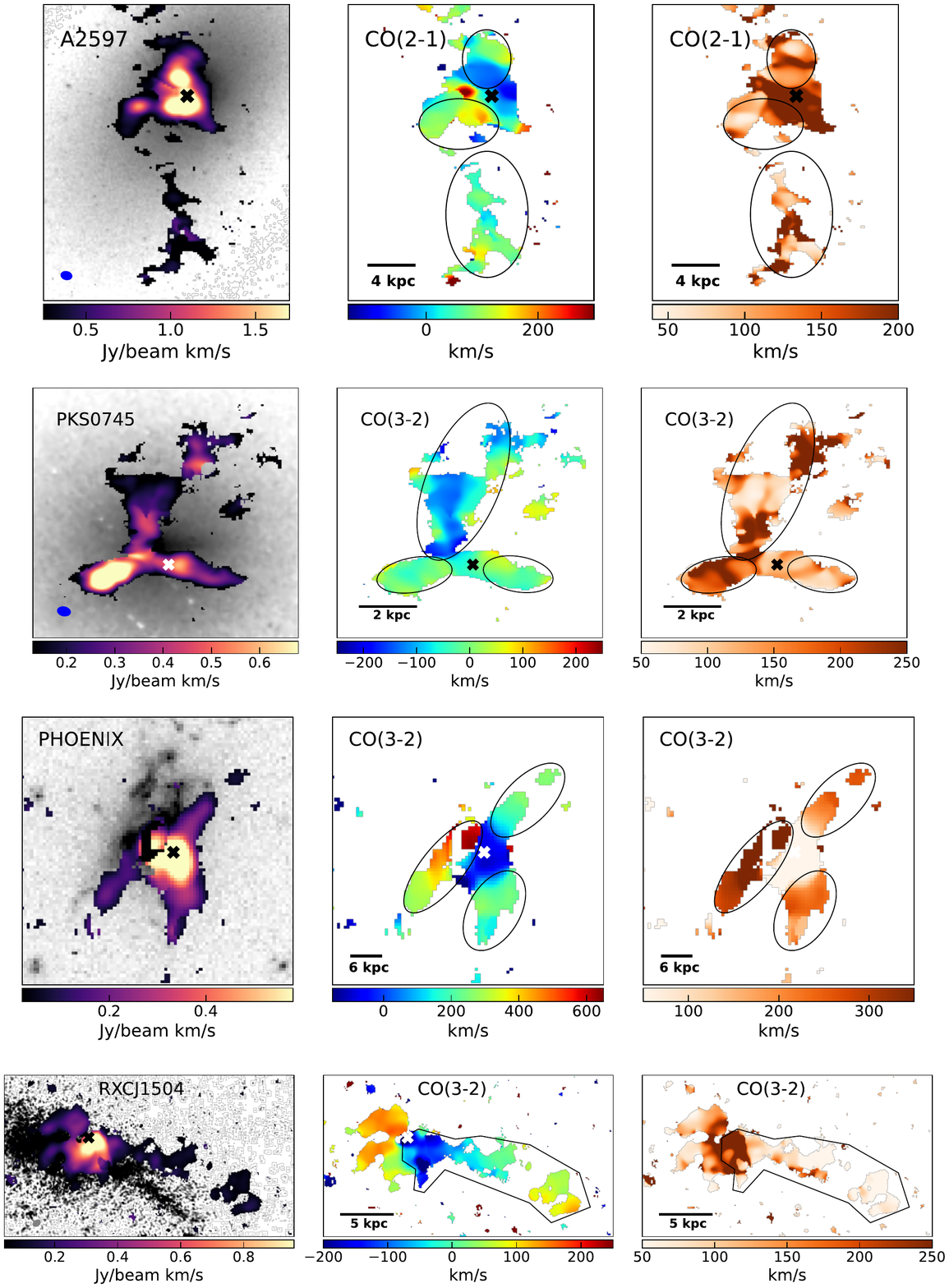}
		\caption{}
	\end{figure*}
	\renewcommand{\thefigure}{A\arabic{figure} (Continued)}
    \addtocounter{figure}{-1}
	\begin{figure*}
	    \centering
		\includegraphics[height=0.9\textheight]{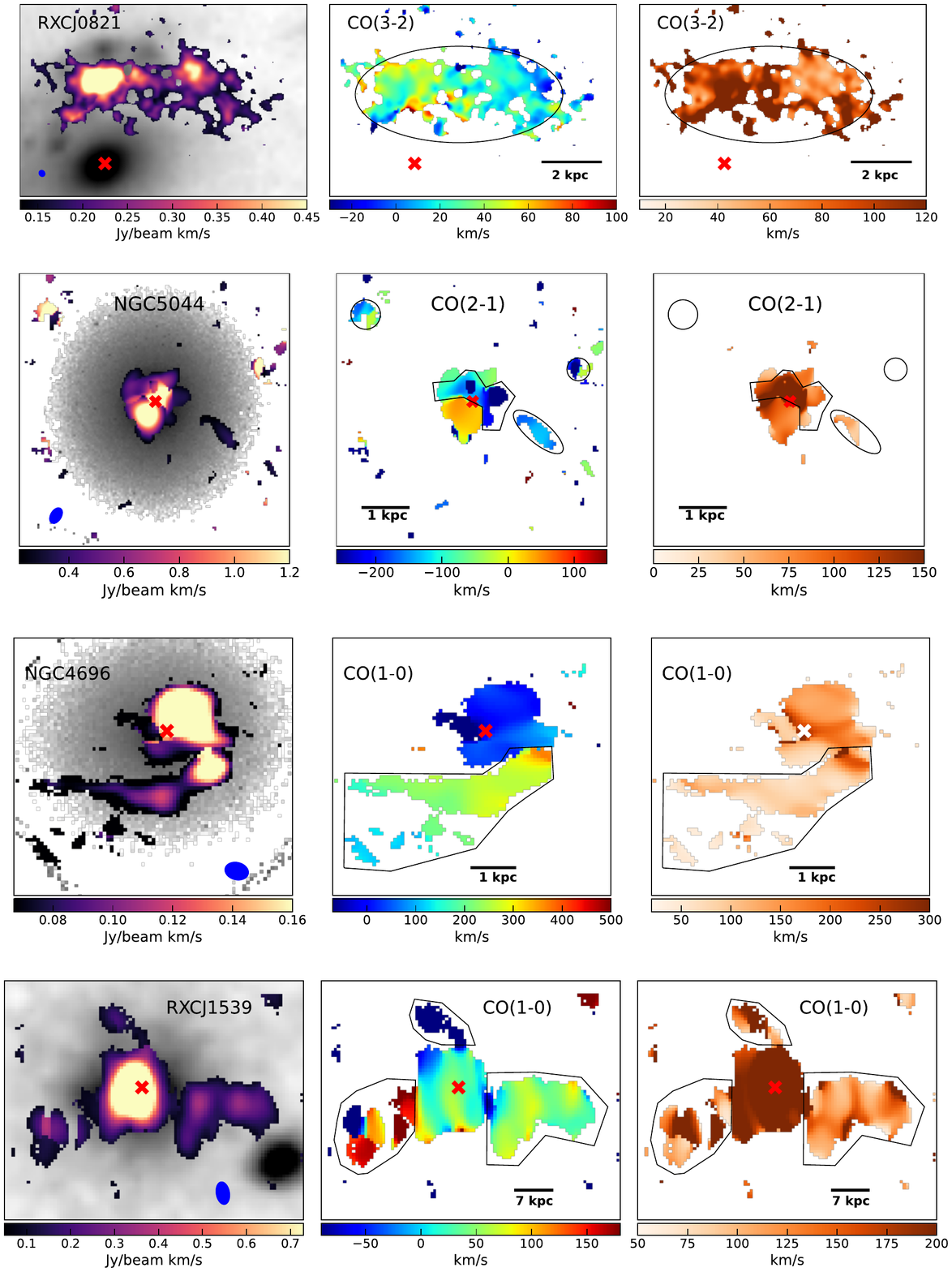}
		\caption{}
	\end{figure*}
	\renewcommand{\thefigure}{A\arabic{figure} (Continued)}
    \addtocounter{figure}{-1}
	\begin{figure*}
		\includegraphics[width=0.9\textwidth]{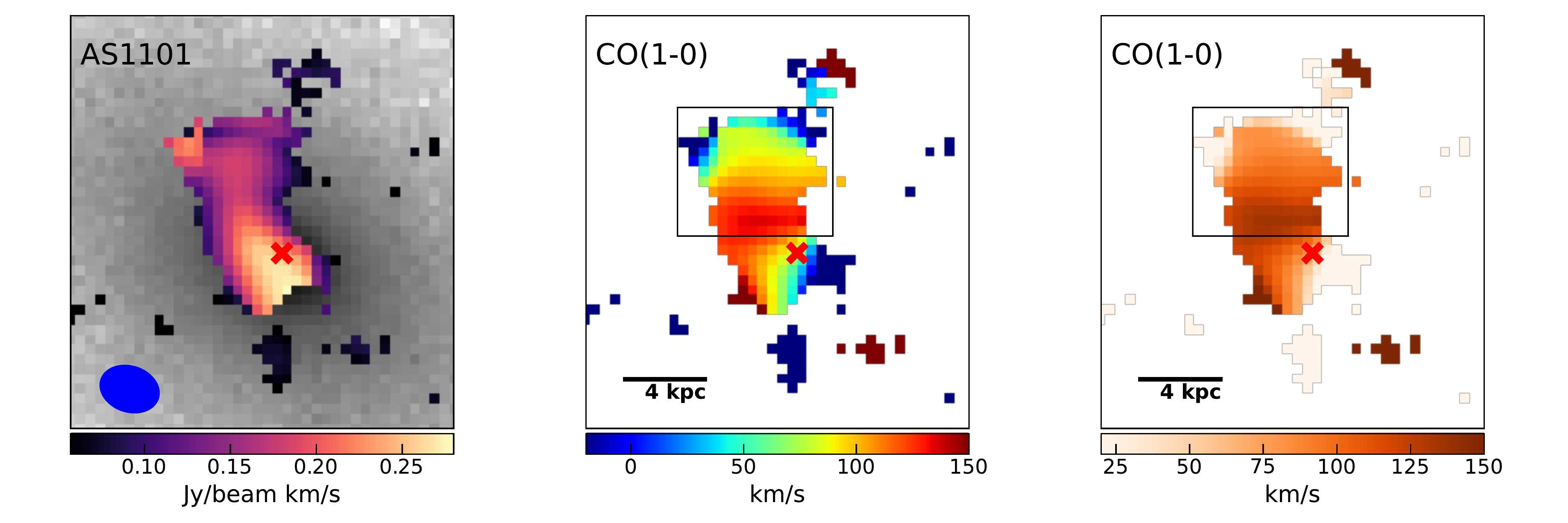}
		\caption{}
	\end{figure*}
	\renewcommand{\thefigure}{A\arabic{figure}}

	\subsection{2A0335+096}
	The molecular gas distribution in 2A0335+096 has a filamentary structure with the filament extending SE to NW of the BCG. There is a large reservoir of molecular gas south of the BCG at a velocity of $-240$ km s$^{-1}$. The north filament appears discontinuous and extends $\sim$ 3 arcseconds (2 kpc) and has an average velocity of 200 km s$^{-1}$. The PV diagram along the BCG does not show a sign of a rotating disk \citep[][see Fig. 11]{vantyghem16}. We consider both the N filament and the southern molecular gas as a flow in this BCG.
	
	\subsection{Abell 1664}
	In Abell 1664, the molecular gas is distributed in a high velocity filament E of the BCG center, a blob to the north of the BCG and another blob coincident with the AGN. The E filament extends up to $\sim$6.37 kpc from the BCG center, whereas the north blob is $\sim$5.8 kpc away from the BCG in projection. The high velocity filament has a velocity of 590 km s$^{-1}$, whereas the blob co-incident with the AGN could be part of a molecular gas disk. We consider the high velocity filament as the flow as shown by elliptical region in Fig~\ref{maps}. \citet[][Fig. 4 and 5]{russell14} show PV diagram along the flow filament and across the BCG. There is a possibility of molecular gas disk across the BCG, the flow filament has a different velocity structure than the central gas reservoir and is not part of the potential molecular gas disk.
	
	\subsection{Abell 1795}

	The molecular gas in A1795 is distributed in two filaments extending to the north and southwest of the BCG centre, and in a central region around the BCG's nucleus. The north filament is curled up around a radio jet (see \cite{russell17b}) with a projected length of 4.2 arcsec (5.1 kpc) and the SW filament is more extended at a projected length of 6 arcsec (7.3 kpc). Both filaments have a distinct velocity structure compared to the central clump, which has a velocity centroid of $\sim$1 km s$^{-1}$ with respect to systemic velocity. The SW filament has a shallower velocity gradient with velocity centroid lying between $-$80 to $-$180 km s$^{-1}$ from the tail-end to the base, respectively. In the N filament, the velocity gradient increases from $-$270 km s$^{-1}$ near the edge to 0 km s$^{-1}$ at the base. The two filaments are considered as flows in our analysis. The PV diagrams presented in \citet{russell17b} show no evidence of a rotating disk. We consider the N and the S filament as flows in our analysis.
	
	\subsection{Abell 1835}

	The molecular gas in Abell 1835 is mostly located within 3 kpc of the BCG centre and peaks within a few tens of km s$^{-1}$ of the galaxy's systemic velocity. However, there are filaments extending $~\sim$12 kpc towards N and SE of the nucleus behind X-ray cavities. The N filament has a narrow component with a velocity centre close to the systemic velocity and a broad component with a velocity centre at $-$230 km s$^{-1}$. The SE filament has an average velocity shift of 40 km s$^{-1}$ and another faint $\sim$200 km s$^{-1}$ component. The PV along the BCG does not show a clear sign of a molecular gas disk (see Fig.~\ref{pv}), although most of the molecular gas in the centre is unresolved. The filaments appear to be part of a bipolar outflow. These regions we consider as flows are shown in Fig.~\ref{maps}.
	
	\subsection{Abell 2597}

	In Abell 2597, most of the molecular gas is distributed in bright filaments draped around radio bubbles close to the nucleus and in a fainter, elongated filament extending $\sim$15 kpc south of the BCG. The brighter filaments are $\sim$6.3 kpc long and contain $\sim$50 per cent of molecular gas. Abell 2597 has a complex velocity structure. Filaments have a shallow constant velocity gradient between velocities 60 to 40 km s$^{-1}$. In the innermost 2 kpc, the gradient changes from $-$120 km s$^{-1}$ to 120 km s$^{-1}$. The system also has absorption features at 240, 275, and 335 km s$^{-1}$~\citep{tremblay16}, which indicates that the gas is moving at fast speeds close to the nucleus. The regions considered as flows are shown in Fig.~\ref{maps}. See \citet{tremblay18} for PV diagrams of the filaments and the central gas structure.
	
	\subsection{PKS0745+091}

	The molecular gas in PKS0745+091 is distributed in three filaments extending north, southwest (SW), and southeast (SE) directions, respectively. The northern filament is most extended at 5.73 kpc followed by SW and SE filaments at projected lengths of 3.6 and 2.7 kpc, respectively. The north and SW filaments appear to be trailing behind an X-ray cavity. The PV diagrams reveal a a velocity gradient from $-$150 to 50 km s$^{-1}$ \citep[see Fig. 8,][]{russell16}. We consider these filaments as flows in this BCG and are shown in Fig.~\ref{maps}.
	
	\subsection{Phoenix}

	Phoenix is one of the most luminous galaxy clusters. The BCG in the Phoenix cluster is massive and has an ongoing star formation at a rate of 610 M$_\odot$ year$^{-1}$. It contains a large amount of molecular gas. The molecular gas in Phoenix is distributed around the centre and in three filaments extending 20-24 kpc to N, SE, and S of the BCG centre. Filaments show a smooth velocity gradient with narrow FWHM ($<$250 km s$^{-1}$), indicating ordered flow motion \citep[also see Fig. 5,][]{russell17a}. The velocity centroids vary in the range between 280 km s$^{-1}$ to 0 km s$^{-1}$ in NW filament and $\sim$250 km s$^{-1}$ to $\sim$570 km s$^{-1}$ in the SE filament and 250 to 50 km s$^{-1}$ in the S filament. These filaments shown in Fig.~\ref{maps} are considered flows in this BCG.
	
	\subsection{RXCJ1504}

	RXCJ 1504 has a central clump and a disturbed, clumpy filament extending radially outward to the W the BCG centre. The projected size of the filament is 18 kpc from the BCG centre, and it contains roughly 30 per cent of the total molecular gas in the system. It has a smooth velocity gradient with velocities between 90 km s$^{-1}$ near the tail to $-$210 km s$^{-1}$ at the base with an FWHM less than 100 km s$^{-1}$ throughout most of the filament. We consider this filament as a flow in this BCG which is shown as a region enclosed by a polygon in Fig.~\ref{maps}.
	
	\subsection{RXCJ0821+0752}
	\label{rxcj0821details}

	The molecular gas in RXCJ0821+0752 is distributed in two major clumps. The first clump, which is $\sim$ 3 kpc north of the BCG centre, contains the majority ($\sim$60\%) of molecular gas. It has a velocity centroid of 24 km s$^{-1}$ compared to the systemic velocity. The secondary clump is located 3 kpc west of the first clump and has a velocity centroid of $\sim$20 km s$^{-1}$. The entire molecular gas distribution is offset from the BCG. The position velocity diagram doesn't show any sign of rotation (see Fig.~\ref{pv}). The region considered as a flow is shown in Fig.~\ref{maps}.
	
	\subsection{NGC 5044}

	Most of the molecular gas in NGC 5044 is distributed around the central galaxy, and some of it is detected as individual clouds of molecular gas NE, NW and E of the BCG. There is a very high-velocity cloud blueshifted to velocities of $\sim -500$ km s$^{-1}$ close to the nucleus. The PV diagram across the BCG doesn't show any sign of a rotating disk. We consider the blueshifted components and the clouds of molecular gas as parts of flow in this galaxy. They are shown in Fig.~\ref{maps}.
	
	\subsection{NGC 4696}
	
	The molecular gas in NGC 4696 is distributed in a large clump roughly cospatial with the BCG nucleus and in a $\sim$4 kpc long curved filament S of the BCG. The central gas reservoir has velocities within 50 km s$^{-1}$ of the BCGs systemic velocity, whereas the filament has a velocity gradient from 120 km s$^{-1}$ in the outer parts to $\sim$350 km s$^{-1}$ in the inner part. The PV diagrams do not show any sign of a rotating disk. This extended filament is considered a flow in NGC 4696.
	
	\subsection{RXCJ1539}
    
\begin{figure*}
        \centering
        \includegraphics[height=0.9\textheight]{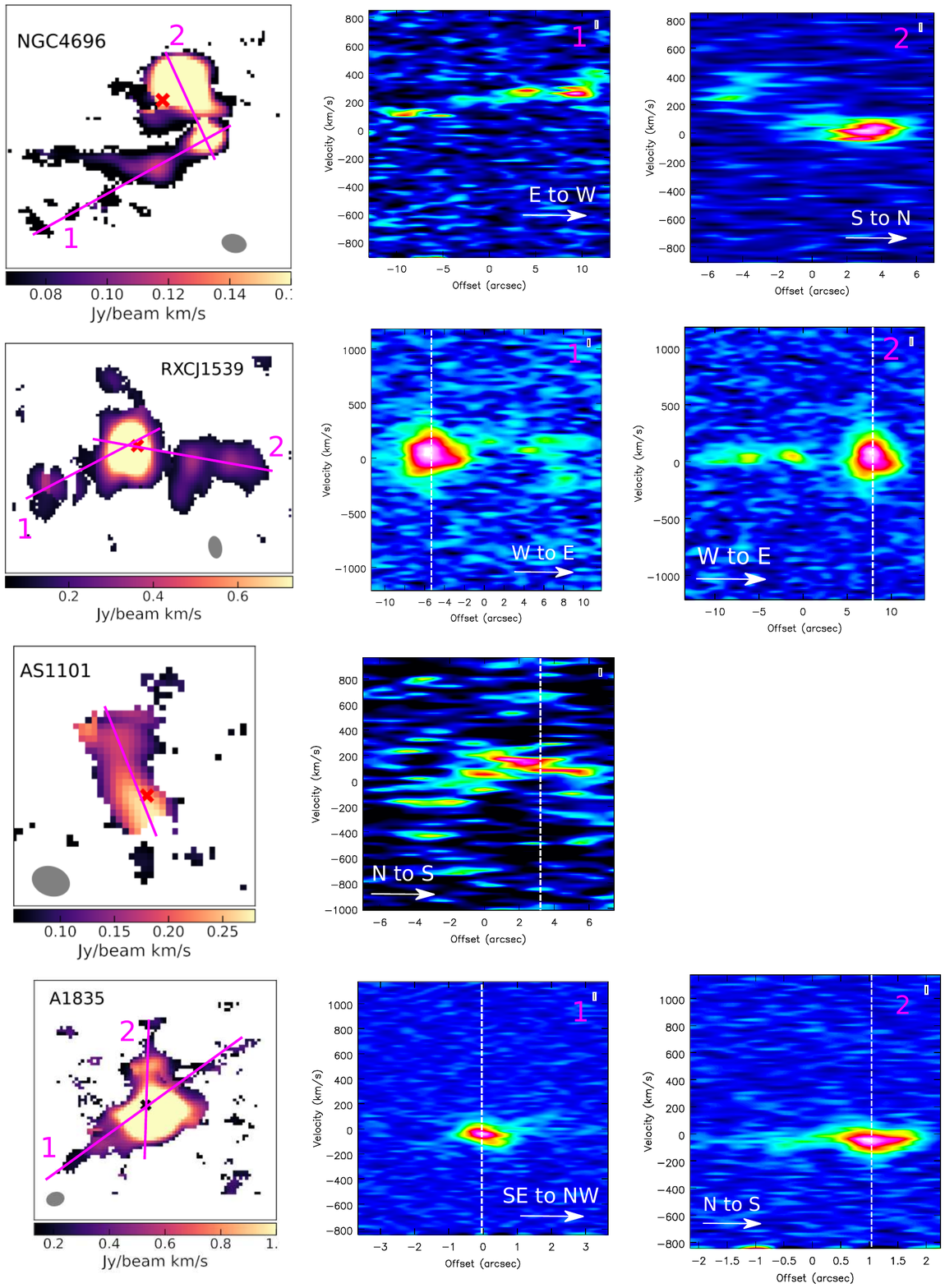}
        \caption{The position-velocity diagrams of BCGs in our sample. The left panel shows an integrated intensity image where the magenta lines indicate the axis used to make PV diagrams. The right panels show the PV diagrams corresponding to the labels shown in the left panel when PV diagrams along multiple axes are shown. The dashed vertical lines show the position of the BCG centre. In all images, the east is to the left and the north is up. The position velocity diagrams do not show any sign of symmetric ordered motion about the nucleus in filaments in any of the BCGs.}
        \label{pv}
    \end{figure*}

    RXCJ1539 has two large filaments to the E and W of the BCG and a central clump of molecular gas co-spatial with the BCG. The E and W filaments extend out to 24 and 27 kpc, respectively. Another small spur of molecular gas is detected to the N of the BCG extending out to 18 kpc. The W filament has a roughly constant velocity of $\sim$50 km s$^{-1}$. The E filament is mostly redshifted to velocities of up to 200 km s$^{-1}$, but also contains a blueshifted component at a velocity of $-$210 km s$^{-1}$. The N filament, however, is entirely blueshifted to velocities of up to $-$285 km s$^{-1}$. The inner 7 kpc radius region around the nucleus has velocities lying between $\sim$0 to 50 km s$^{-1}$. It is no unambiguous evidence of smooth gradient or ordered motion indicating a rotating disk in the PV diagrams shown in Fig.~\ref{pv}, as the central reservoir is unresolved. The three regions we consider as a flow in this BCG are shown in Fig.~\ref{maps}.
    
    \subsection{AS1101}
    
    The molecular gas in AS1101 appears as a single extended filament to the N of the BCG. The filament has a smooth velocity gradient from 140 km s${-1}$ close to the BCG to 40 km s$^{-1}$ in the outer parts. It is extended out up to 8.23 kpc N of the BCG. Some blueshifted emission is detected to the W of the BCG at velocities of $-$130 km s$^{-1}$. The PV diagrams show a smooth velocity gradient across the filament. The filament is indistinguishable from the circumnuclear gas in this data. The region we consider as a flow is shown in Fig.~\ref{maps}.
    
    \renewcommand{\thefigure}{A\arabic{figure} (Continued)}
    \addtocounter{figure}{-1}
	\begin{figure*}
	    \centering
		\includegraphics[width=0.9\textwidth]{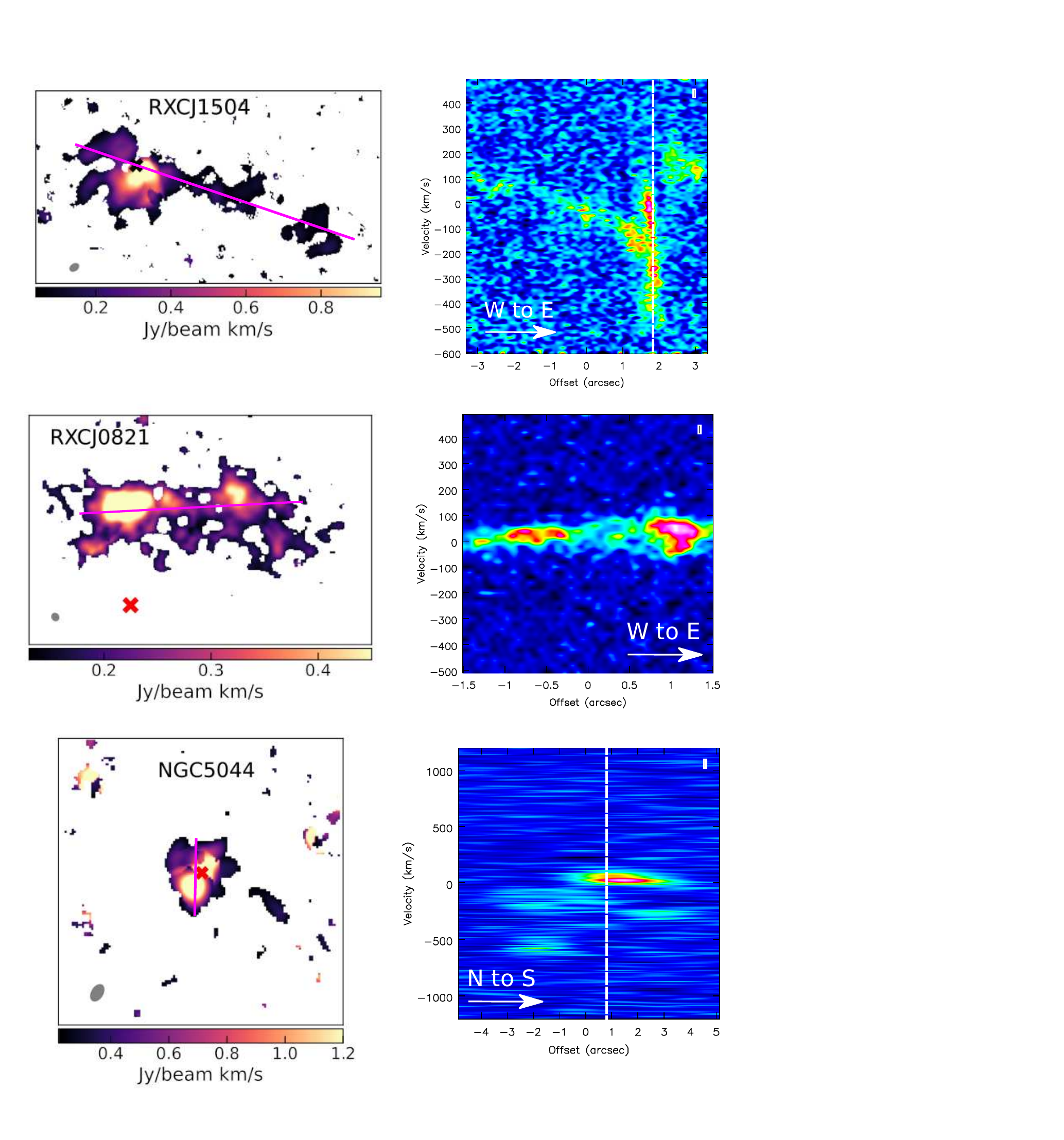}
		\caption{}
	\end{figure*}


\section{Flow properties with adjusted velocities for Fluetsch galaxies}
\label{appB}
In Fig.~\ref{v_comp}, we showed that applying the Fluetsch method to BCGs overestimates flow velocities by a factor of two. Assuming that a similar systematic difference in velocities applies to galaxies in the Fluetsch sample, we divide their outflow velocities by two to estimate new velocities. It results in a reduction of molecular outflow rates by half, kinetic energies of outflows by a factor of 8 and momenta by a factor of 4. Figure~\ref{newplots} shows all the relevant figures where outflow properties for galaxies in the Fluetsch sample are calculated using new velocities. It does not affect correlation coefficients for Fluetsch outflows. The average flow velocities of BCGs lie only 30\% below average flow velocity in AGN hosting galaxies and $~\sim20$\% faster than flow velocities in starburst galaxies. The flow momenta for AGN, composite and starburst galaxies are 3.3$\times$10$^{49}$, 5.2$\times$10$^{48}$ and 8.6$\times$10$^{48}$ g cm s$^{-1}$, respectively. They are 10--15 times smaller than flow momenta in BCGs. The loading factors do not change significantly. More galaxies lie within theoretical maximum limits for energy and momentum conserving flows. At the same time, the ratio of kinetic power of the flow to the power of the driving mechanism falls below 0.05\% for some galaxies, indicating a weak coupling. The lifting factor remains unaffected, as it does not depend on the flow velocity. Thus, qualitatively, these results are consistent with the results presented in the sections above.

\renewcommand{\thefigure}{A\arabic{figure}}
\addtocounter{figure}{2}
\begin{figure*}
    \centering
    \includegraphics[width=0.3\textwidth]{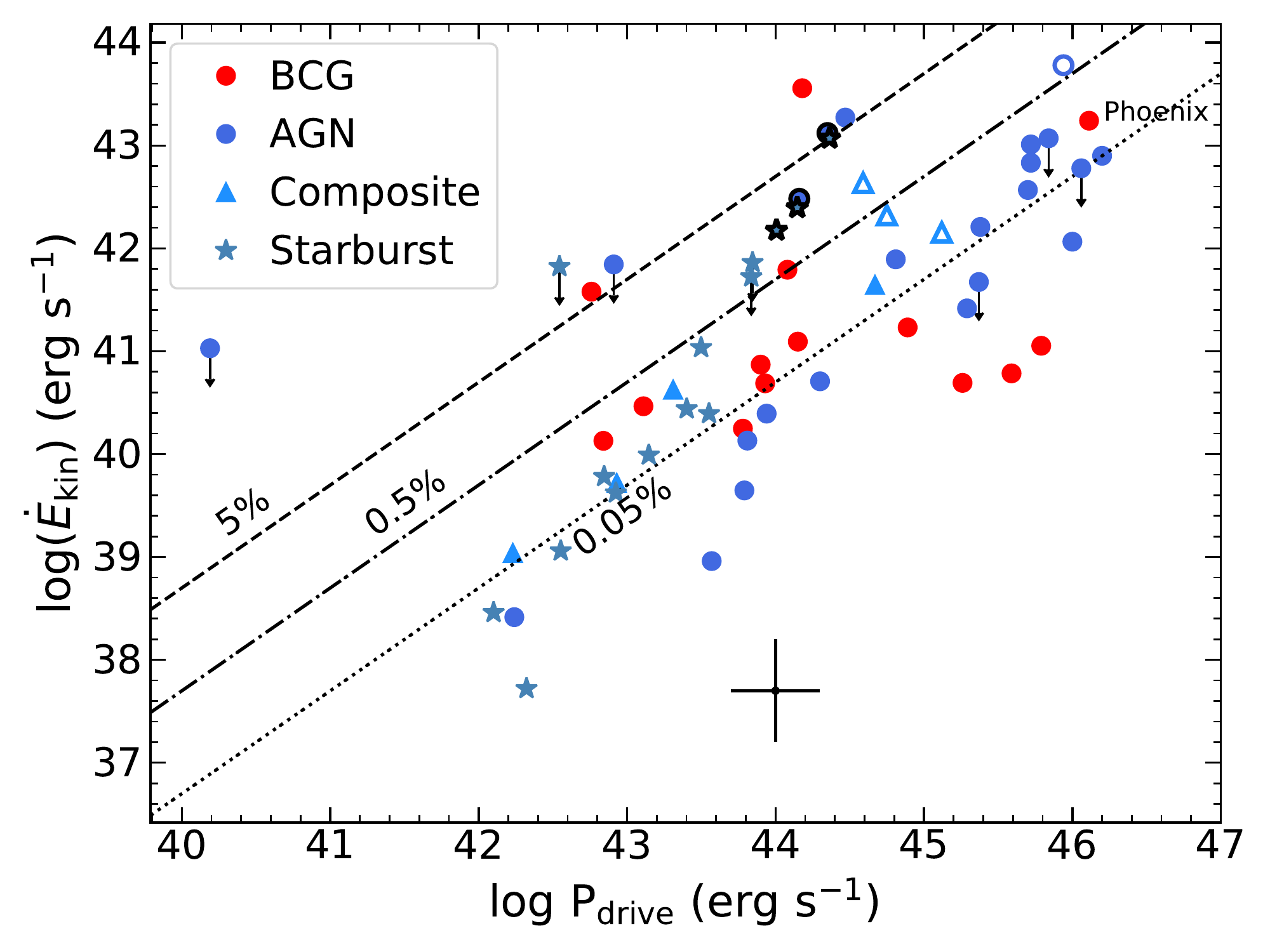}
    \includegraphics[width=0.3\textwidth]{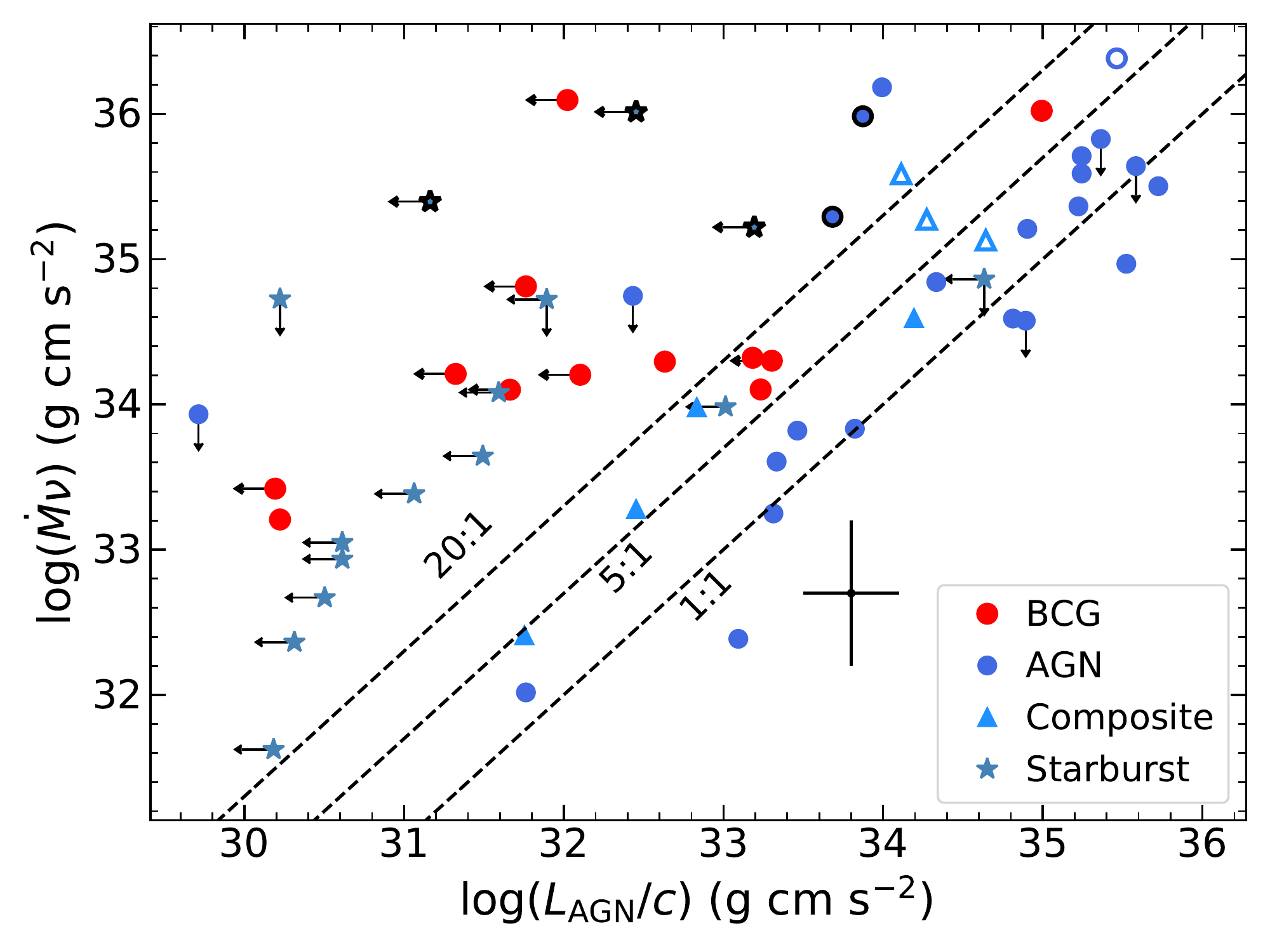}
    \includegraphics[width=0.3\textwidth]{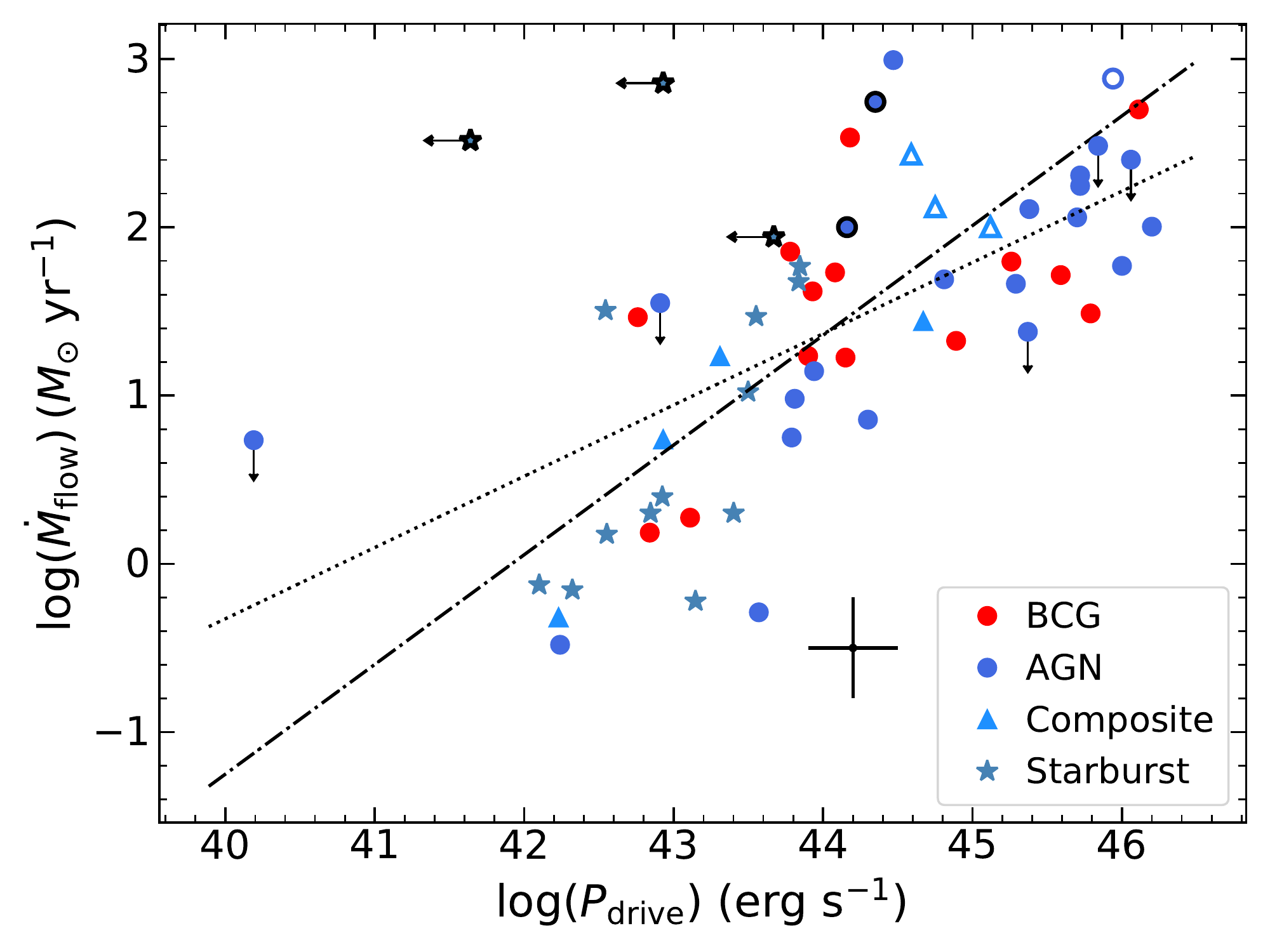}
    \newline
    \includegraphics[width=0.3\textwidth]{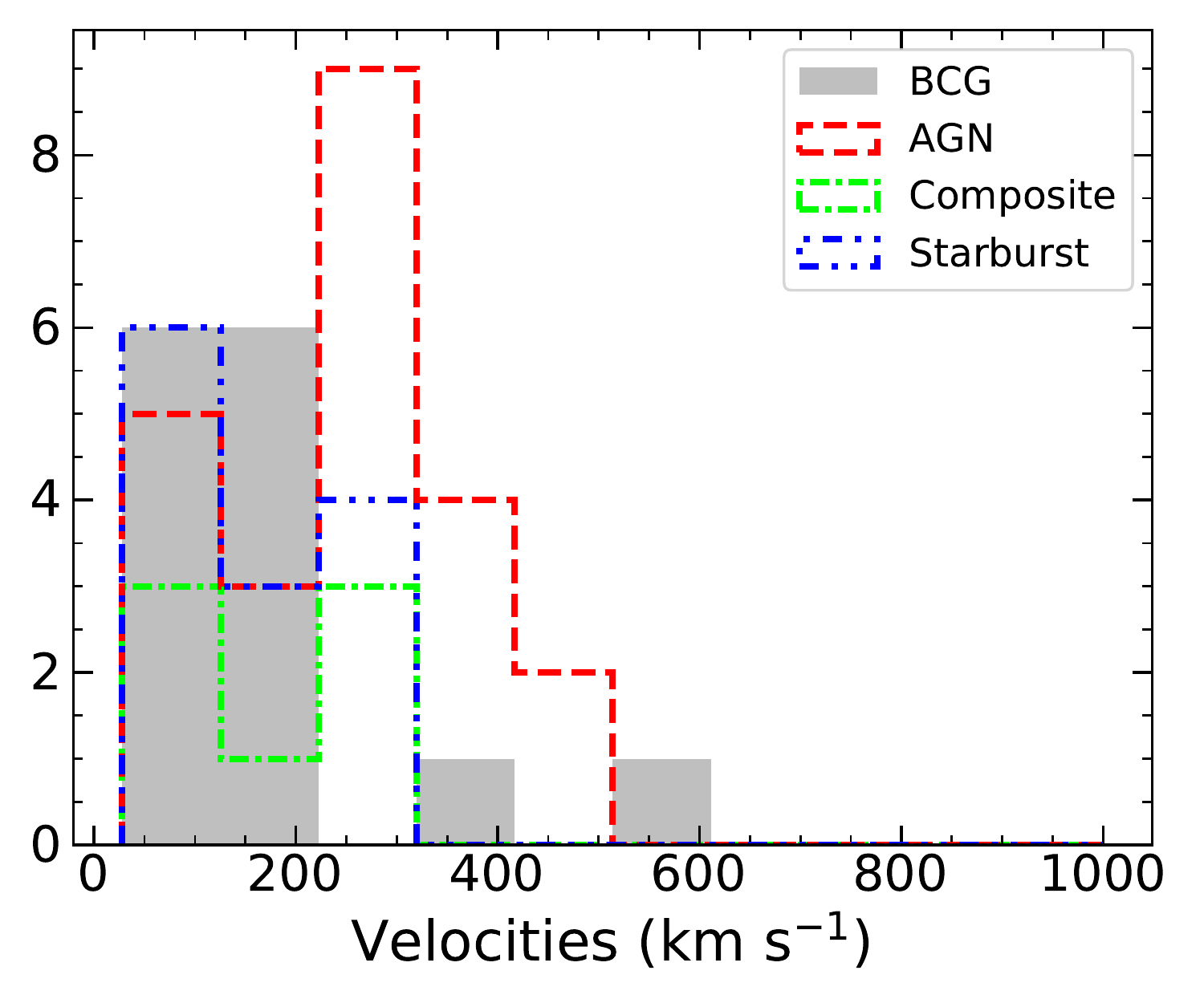}
    \includegraphics[width=0.3\textwidth]{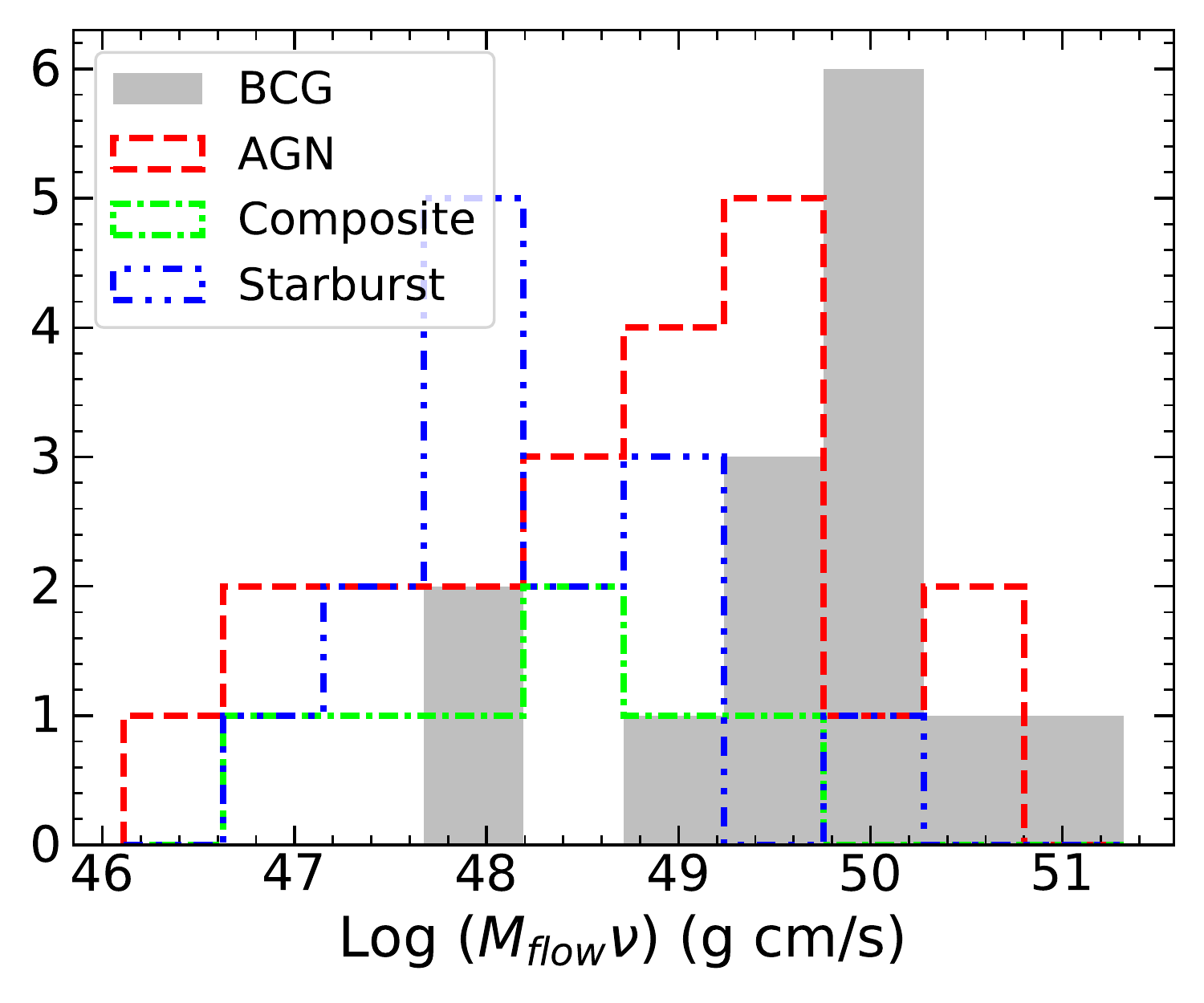}
    \includegraphics[width=0.3\textwidth]{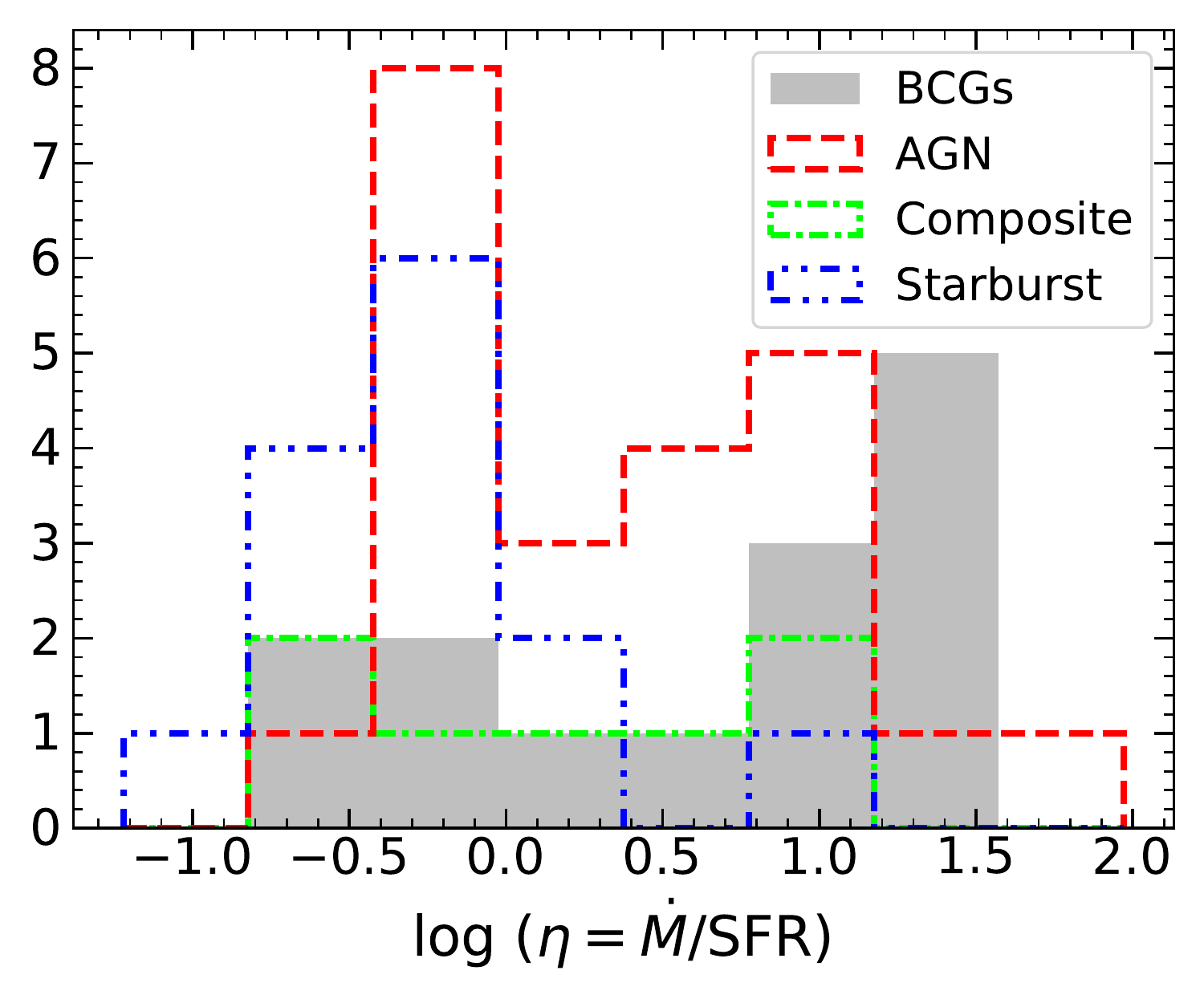}
    \caption{We show figures~\ref{power},~\ref{forces},~\ref{M-P},~\ref{velhist},~\ref{mom-hist},~\ref{loadingfact} from top left to bottom right panels, respectively, with the molecular flow rate, momentum, kinetic energy of the flow calculated with adjusted flow velocities for fluetsch galaxies as described in Appendix~\ref{appB}.}
    \label{newplots}
\end{figure*}

\bsp	
\label{lastpage}
\end{document}